\documentclass[11pt]{article}
\usepackage{verbatim}
\usepackage{amssymb,amsfonts,amsthm,amsmath,amssymb,amscd,amstext,epsfig}
\usepackage{color}
\usepackage{hyperref}
\definecolor{darkgreen}{rgb}{0,.5,0}

\def\sgn{\operatorname{sgn}}
\def\Vol{\operatorname{Vol}}

\def\Tr{\operatorname{Tr}}
\def\calA{{{\mathcal A}}}
\def\calF{{{\mathcal F}}}
\def\calN{{{\mathcal N}}}
\def\calR{{{\mathcal R}}}
\def\calW{{{\mathcal W}}}
\def\calJ{{{\mathcal J}}}
\def\calI{{{\mathcal I}}}
\def\d{{\rm d}}
\def\be{\begin{equation}}
\def\ee{\end{equation}}
\def\CI{{\cal I}}

\def\CN{{\cal N}}

\def\BZ{\mathbb{Z}}

\def\BC{\mathbb{C}}

\newcommand{\Om}[1]{\text{O}{#1}^-}
\newcommand{\tOm}[1]{\widetilde{\text{O}{#1}}^-}
\newcommand{\Op}[1]{\text{O}{#1}^+}
\newcommand{\tOp}[1]{\widetilde{\text{O}{#1}}^+}
\newcommand{\Opm}[1]{\text{O}{#1}^\pm}
\newcommand{\tOpm}[1]{\widetilde{\text{O}{#1}}^\pm}

\newcommand{\half}{{1\over 2}}

\begin{document}

\thispagestyle{empty}

\begin{flushright}
PUPT-2404\\
YITP-SB-12-02\\

%arXiv:yymm.nnnn [hep-th]
\end{flushright}

\begin{center}
\vspace{1cm} { \LARGE {\bf The ABCDEF's of Matrix Models for \\
\vspace{0.3cm}
 Supersymmetric Chern-Simons Theories}}

\vspace{1.1cm}

Daniel R.~Gulotta$^*$, Christopher P.~Herzog$^{*\dagger}$,
and Tatsuma Nishioka$^*$

\vspace{0.7cm}

{$*$ Department of Physics, Princeton University \\
     Princeton, NJ 08544, USA }

 \vspace{0.2cm}
 
 {$\dagger$ YITP, Stony Brook University \\
 Stony Brook, NY 11794, USA }

\vspace{0.7cm}

%{\tt dgulotta, cpherzog, nishioka@princeton.edu} \\

\vspace{1.5cm}

\end{center}

\begin{abstract}
\noindent
We consider $\CN =3$ supersymmetric Chern-Simons gauge
theories with product unitary and orthosymplectic groups and bifundamental and fundamental fields.
We study the partition functions on an $S^3$ by using the Kapustin-Willett-Yaakov matrix model. The saddlepoint equations in a large $N$ limit lead to
a constraint that the long range forces between the eigenvalues must cancel;   
the resulting quiver theories are of affine Dynkin type.
We introduce a folding/unfolding trick which lets us, at the level of the large $N$ 
matrix model, (i) map quivers with orthosymplectic groups to those with unitary groups, and (ii) obtain non-simply laced quivers from the 
corresponding simply laced quivers using a $\BZ_2$ outer automorphism.
The brane configurations of the quivers are described in string theory and 
the folding/unfolding is interpreted as the addition/subtraction of orientifold and orbifold planes.
We also relate the $U(N)$ quiver theories to the affine ADE quiver matrix models
with a Stieltjes-Wigert type potential, and derive the generalized Seiberg duality 
in $2+1$ dimensions from Seiberg duality in $3+1$ dimensions.
\end{abstract}

\pagebreak

\tableofcontents

\setcounter{page}{1}
\setcounter{equation}{0}

%%%%%%%%%%%%%%%%%%%%%%%%%%%%%%%%%%%%
\section{Introduction}
%%%%%%%%%%%%%%%%%%%%%%%%%%%%%%%%%%%%

In this paper, we continue an investigation, 
started in refs.\ \cite{Herzog:2010hf,Gulotta:2011aa,Gulotta:2011si,Gulotta:2011vp},
 of the
 large $N$ limit of the $S^3$ partition function 
of supersymmetric (SUSY) Chern-Simons (CS) theories.
The partition function $Z_{S^3}$  is calculated using the matrix model derived in ref.\ \cite{Kapustin:2009kz} by localization (later improved by \cite{Jafferis:2010un,Hama:2010av} to allow matter fields to acquire anomalous dimensions).  
For the CS theory at its superconformal fixed point, the matrix model of ref.\ \cite{Kapustin:2009kz} computes exactly the partition function and certain supersymmetric Wilson loop expectation values.  
The theories we examine here have ${\mathcal N}=3$ SUSY, a product classical gauge group structure, and bifundamental field content summarized by a quiver diagram.
While in our previous work  \cite{Herzog:2010hf,Gulotta:2011aa,Gulotta:2011si,Gulotta:2011vp} we examined CS theories that had only unitary groups, in this work we allow for $O(N)$ and $USp(2N)$ groups as well.

We are motivated by the hope that SUSY gauge theories in 2+1 dimensions 
will help us learn about general features of 2+1 dimensional gauge theories which in turn might shed light on certain condensed matter systems with emergent gauge symmetry at low temperatures.  
One powerful tool for examining these 2+1 dimensional SUSY CS matter theories is the AdS/CFT correspondence \cite{Maldacena:1997re,Gubser:1998bc,Witten:1998qj}.
In this 2+1 dimensional ${\cal N}=3$ SUSY context, 
the correspondence can be motivated by placing a stack of $N$ M2 branes at 
the singularity of a four complex dimensional hyperk\"ahler cone.  On the one hand, the low 
energy description of the M2-branes is the CS matter theory.  An $N$ fold symmetric product of the cone is a branch of the moduli space.  On the other, there is a dual eleven dimensional supergravity description of the theory: Close to the M2 branes the geometry is $AdS_4 \times Y$ where $Y$ is a seven real dimensional base of the cone (a tri-Sasaki Einstein manifold) that is threaded by $N$ units of $\star F_4$ flux.  In the limit $N \to \infty$, 
the correspondence maps the CS matter theory in a strong coupling limit to this classical supergravity description where correlation functions can be easily computed.

In order to use AdS/CFT as a tool to deduce universal properties of 
strongly interacting gauge theories, 
one should first understand what kinds of strongly 
interacting gauge theories have classical gravity duals.  
In this paper we use the matrix model to find a large class of ${\cal N}=3$ SUSY CS matter theories with a dual eleven dimensional supergravity description.  
These theories are described by affine Dynkin diagrams where the nodes are the classical groups $U(N)$, $O(N)$, and $USp(2N)$ and the arrows are bifundamental fields.   To each group factor, we associate a CS level. 

On the matrix model side, the existence of an $AdS_4 \times Y$ eleven dimensional supergravity limit appears to be related to the cancellation of the long range forces between the eigenvalues in a saddle point approximation along with a constraint on the sum of the CS levels 
\cite{Herzog:2010hf, Jafferis:2011zi}.
 Given these two conditions, which we describe in more detail in section \ref{ss:unfold}, the free energy, defined as 
\be
F \equiv - \ln Z_{S^3} \ , 
\ee
will scale as $F \sim N^{3/2}$.  This scaling is then in agreement with an older gravity calculation \cite{Emparan:1999pm}, 
\be
\label{emparanresult}
F = \frac{\pi L^2}{2 G_N} \ ,
\ee
where $L$ is the radius of curvature of the $AdS_4$ and $G_N$ is an effective four-dimensional Newton constant.\footnote{%
The $N^{3/2}$ scaling was seen earlier in the thermal free energy \cite{Klebanov:1996un}. 
}
The quantization of $L$ in Planck units implies that at large $N$ eq.\ (\ref{emparanresult}) becomes
\cite{Herzog:2010hf}
\begin{align}\label{GRfree}
	F = N^{3/2}\sqrt{\frac{2\pi^6}{27\operatorname{Vol}(Y)}} \ .
\end{align}
(Here, the volume of $Y$ is computed with an Einstein metric that satisfies the
normalization condition $R_{mn} = 6 g_{mn}$.)\footnote{%
There is a more nuanced story relating the cancellation of long range forces to the existence of a gravity dual. The cancellation we discuss here depends on an ansatz for the eigenvalue distribution.  
 A more elaborate ansatz or a different large $N$ approximation of the matrix model may also
 lead to an $N^{3/2}$ scaling.
  For example, while for $\calN=2$ CS theories with chiral bifundamental fields the long range forces do not naively cancel, ref.\ \cite{Amariti:2011uw} proposed a remedy that involves first symmetrizing the matrix model integrand with respect to a $\mathbb{Z}_2$ subgroup.  
  On the other hand, relaxing the constraint on the CS levels should lead to type IIA supergravity duals \cite{Jafferis:2011zi}.
  }

\begin{figure}[h]
\begin{align}
A_n^{(1)} &&& \parbox{5cm}{\quad\scalebox{0.3}{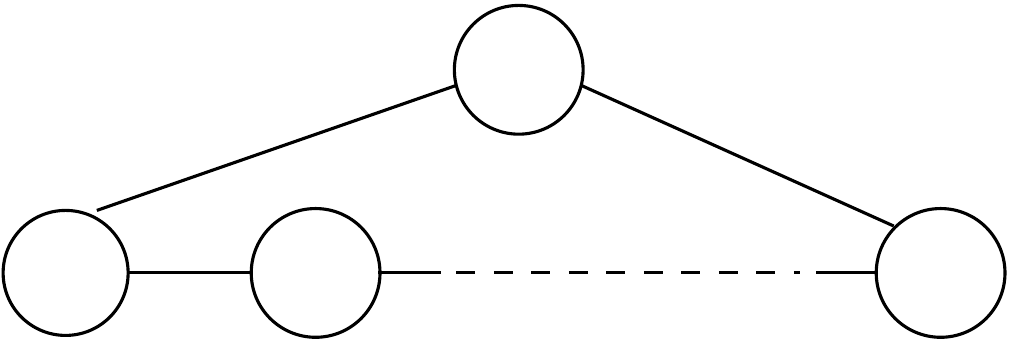}} \notag \\
D_n^{(1)} &&& \parbox{5cm}{\scalebox{0.3}{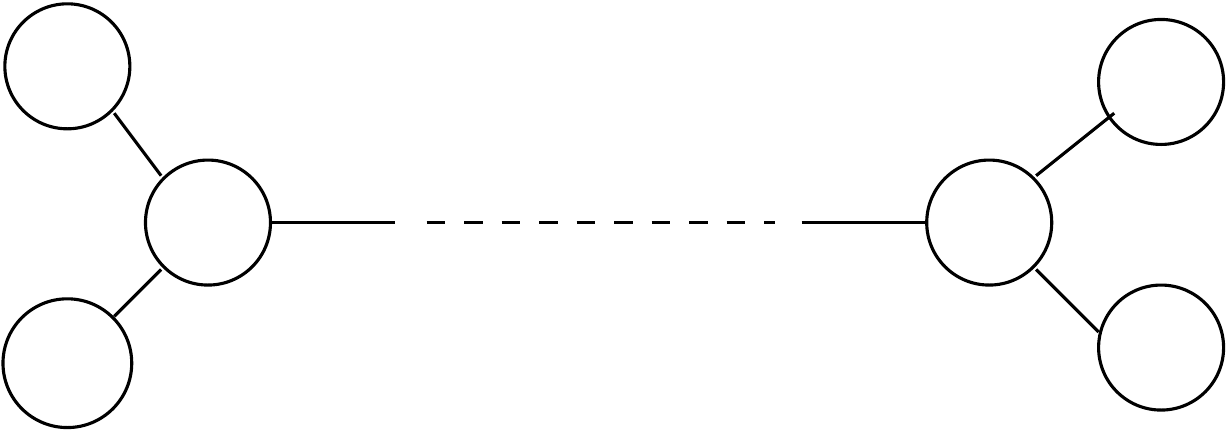}} \notag \\
E_6^{(1)} &&& \parbox{5cm}{\scalebox{0.3}{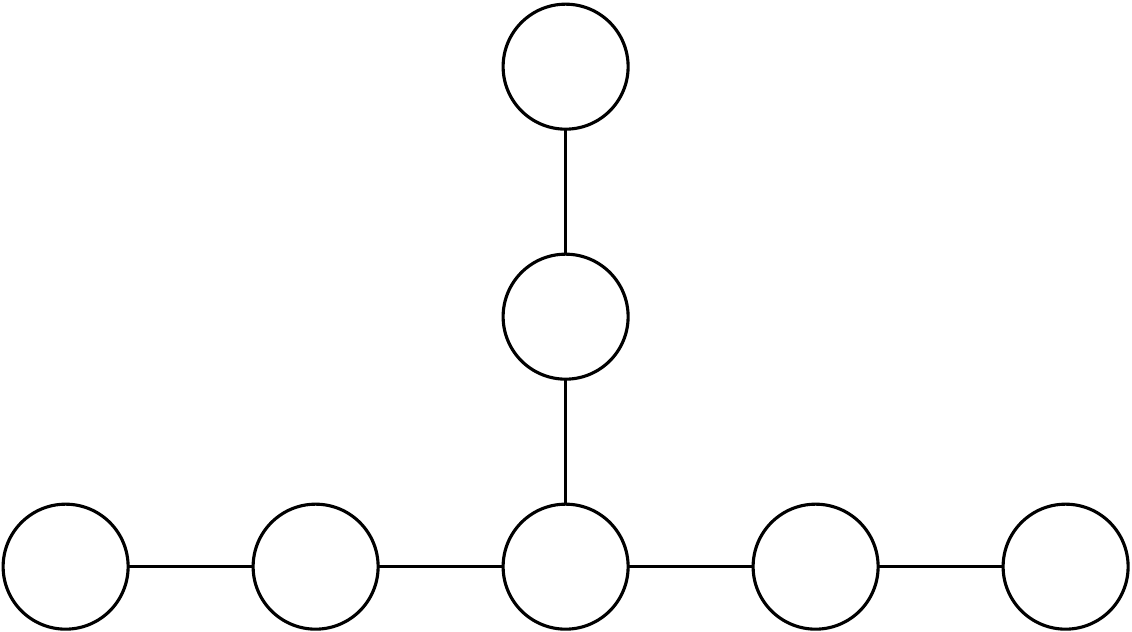}} \notag \\
E_7^{(1)} &&& \parbox{5cm}{\scalebox{0.3}{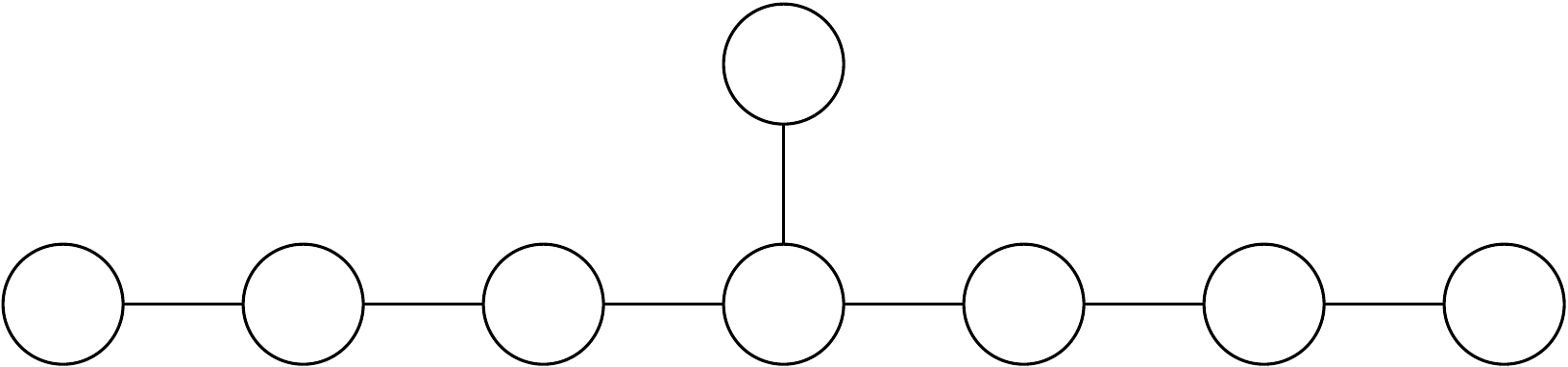}} \notag \\
E_8^{(1)} &&& \parbox{5cm}{\scalebox{0.3}{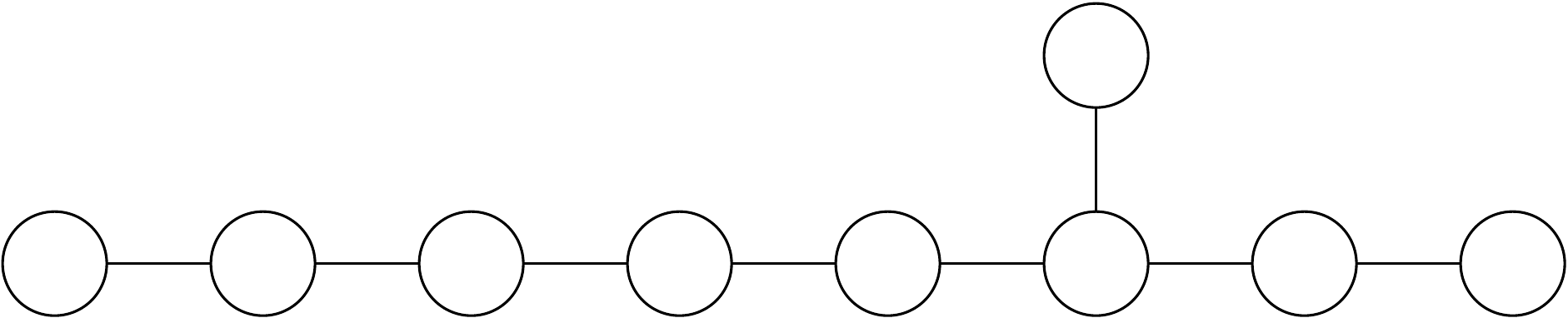}} \notag
\end{align}
\caption{
\label{fig:ADEquivers}
Affine ADE quivers. Each node corresponds to a gauge group, and each edge
corresponds to a bifundamental hypermultiplet. When two adjacent nodes
are $O\times USp$ or $USp \times O$, the edge between them means
a half hypermultiplet.
The numbers in the circles denote the comarks or dual Kac labels \cite{Kac:1990gs,DiFrancesco:1997nk}.  The numbers are also the ranks of the gauge groups divided by an overall factor of $N$.}
\end{figure}

\begin{figure}[h]
\begin{align}
A^{(2)}_{2n-1} &&& \parbox{5cm}{\scalebox{0.3}{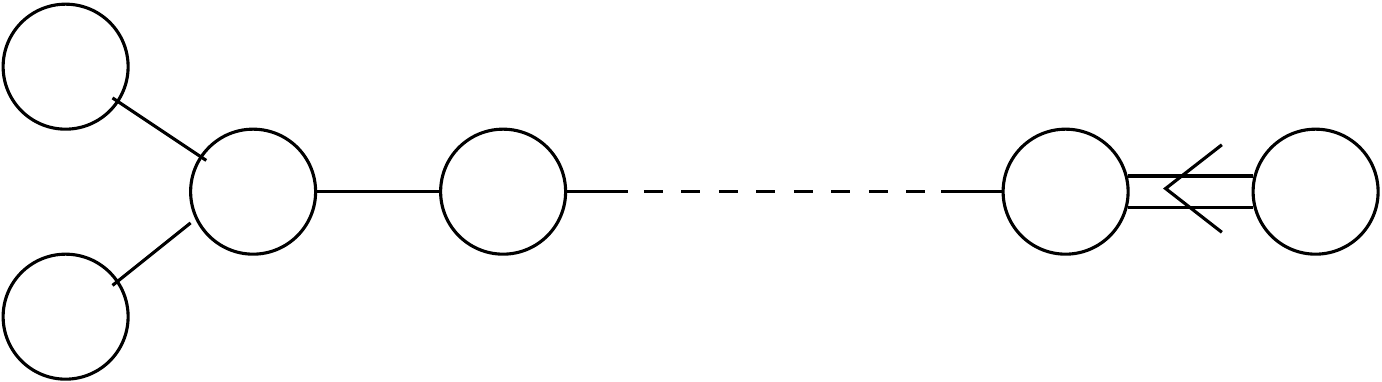}}
&& D_{2n}^{(1)} & \parbox{5cm}{\scalebox{0.3}{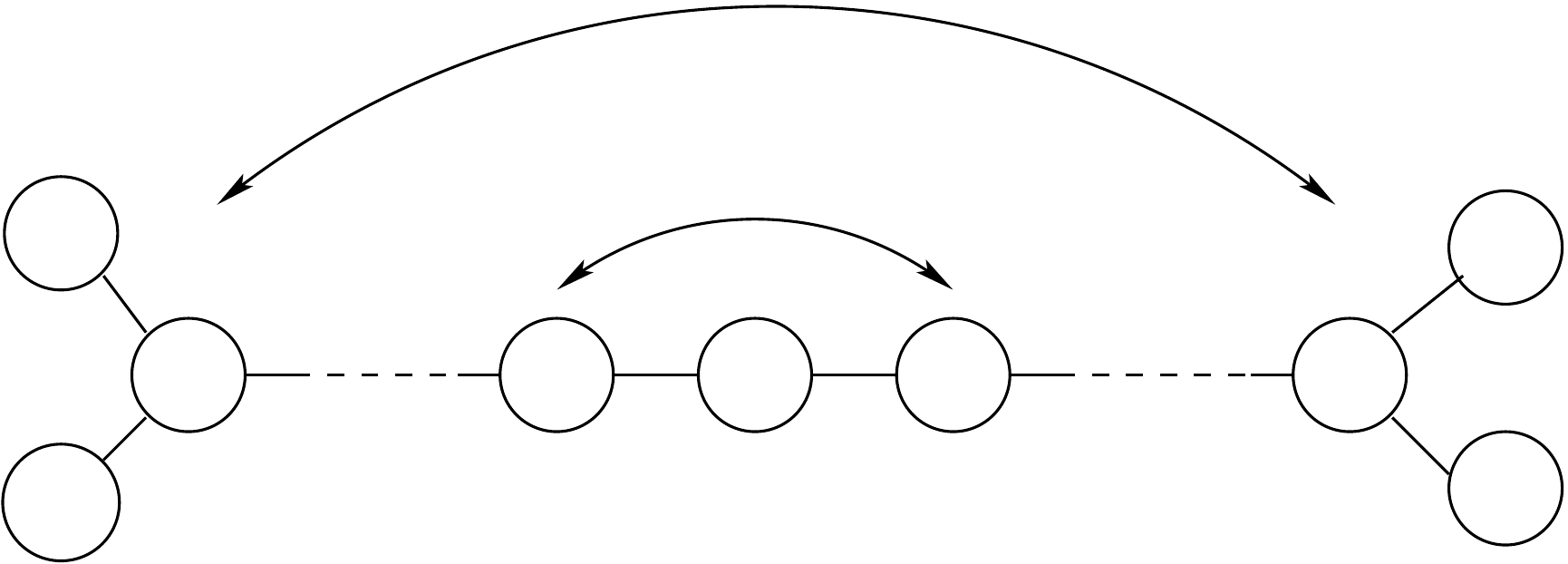}} \notag \\
B_n^{(1)} &&& \parbox{5cm}{\scalebox{0.3}{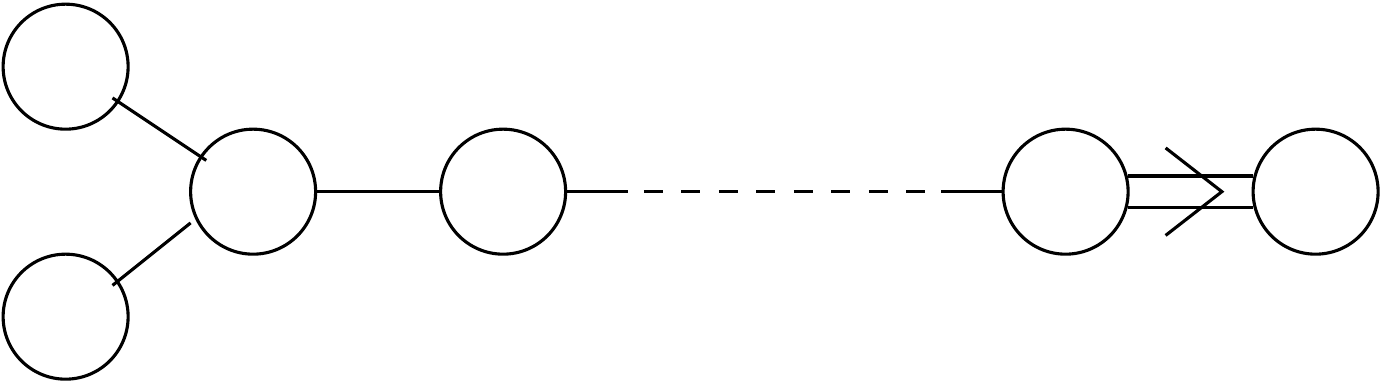}}
&&  D_{n+1}^{(1)} & \parbox{5cm}{\scalebox{0.3}{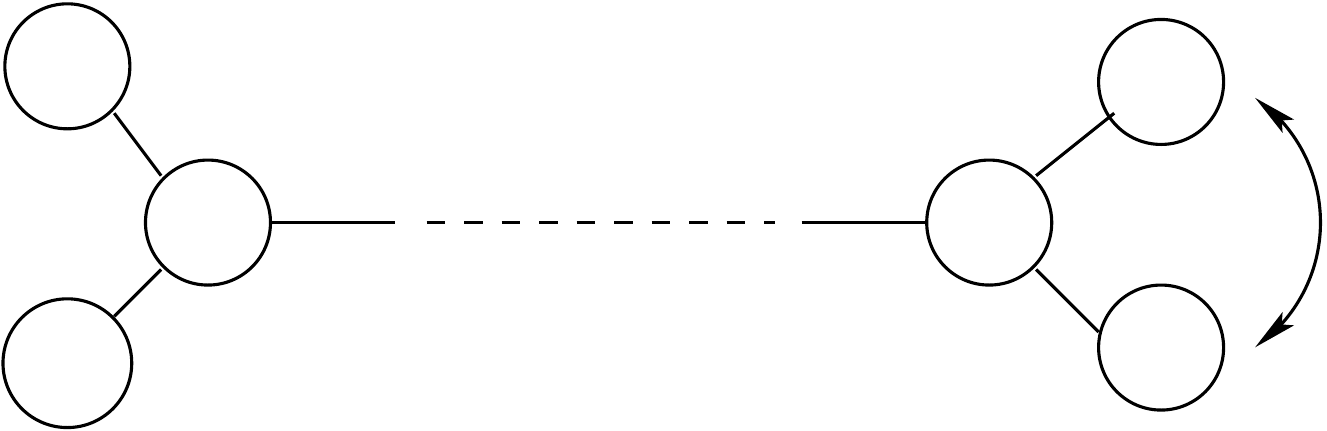}}\notag \\
C_n^{(1)} &&& \parbox{5cm}{\scalebox{0.3}{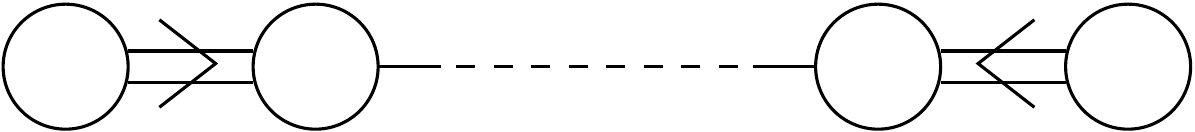}}
&&  A_{2n-1}^{(1)} & \parbox{5cm}{\scalebox{0.3}{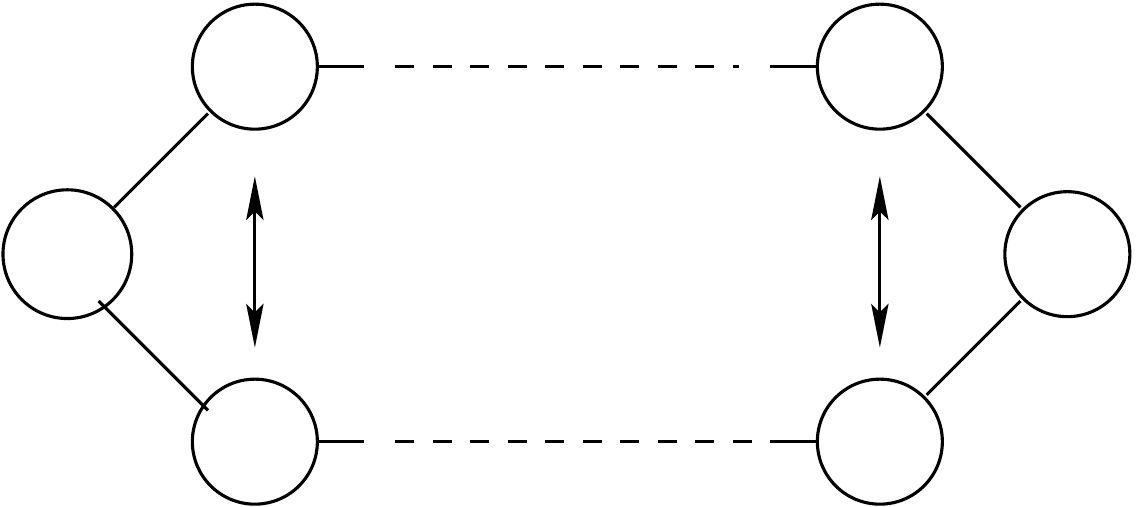}}\notag \\
D^{(2)}_{n+1} &&& \parbox{5cm}{\scalebox{0.3}{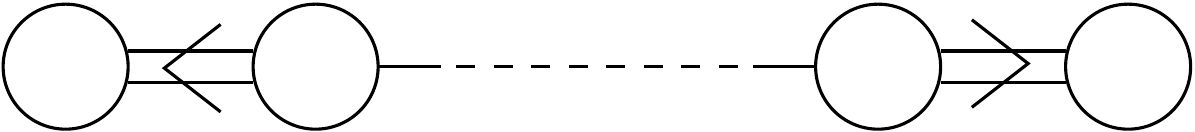}}
&& D_{n}^{(1)} & \parbox{5cm}{\scalebox{0.3}{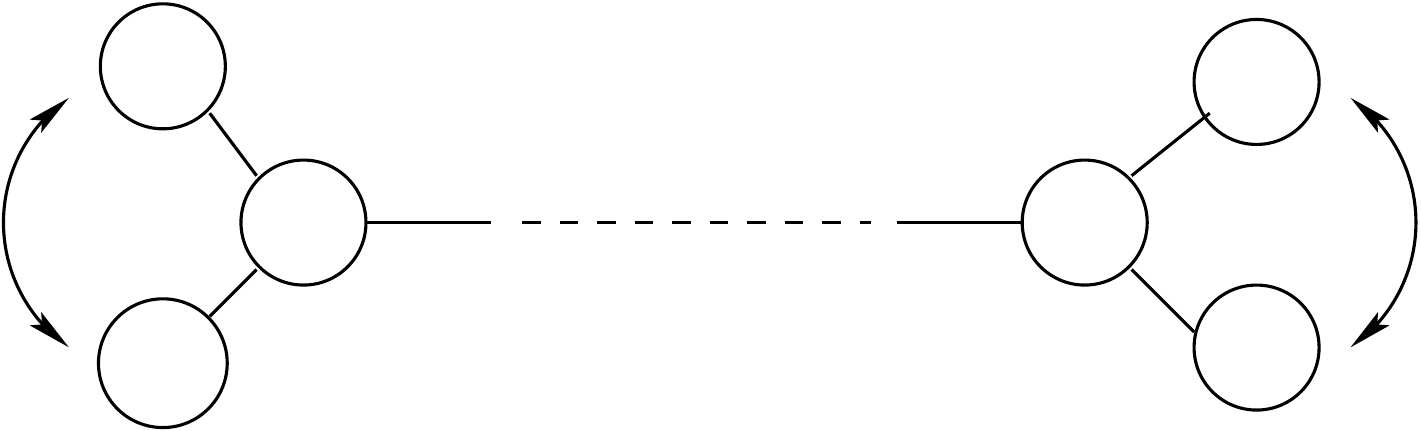}} \notag \\
E^{(2)}_6 &&& \parbox{5cm}{\scalebox{0.3}{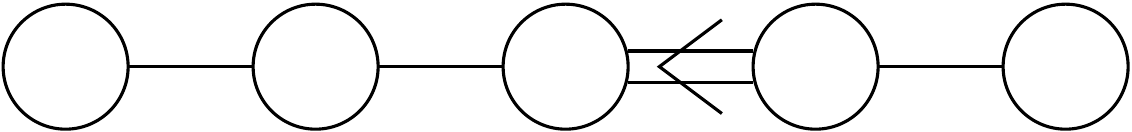}}
&& E_7^{(1)} & \parbox{5cm}{\scalebox{0.3}{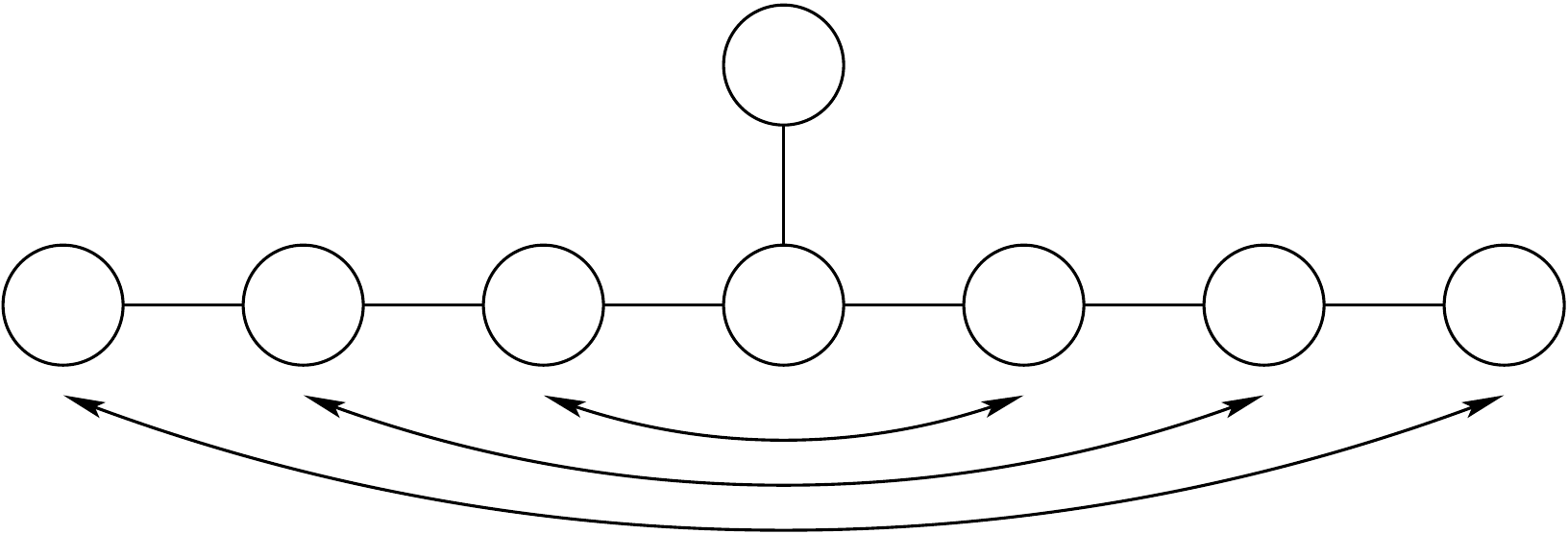}}\notag \\
F_4^{(1)} &&& \parbox{5cm}{\scalebox{0.3}{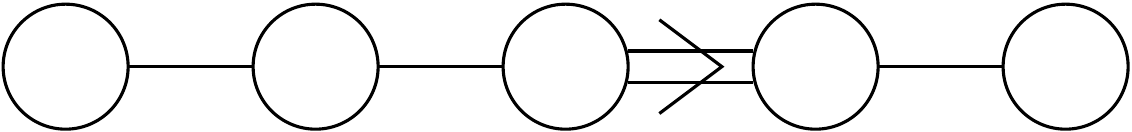}}
&& E_6^{(1)} & \parbox{5cm}{\scalebox{0.3}{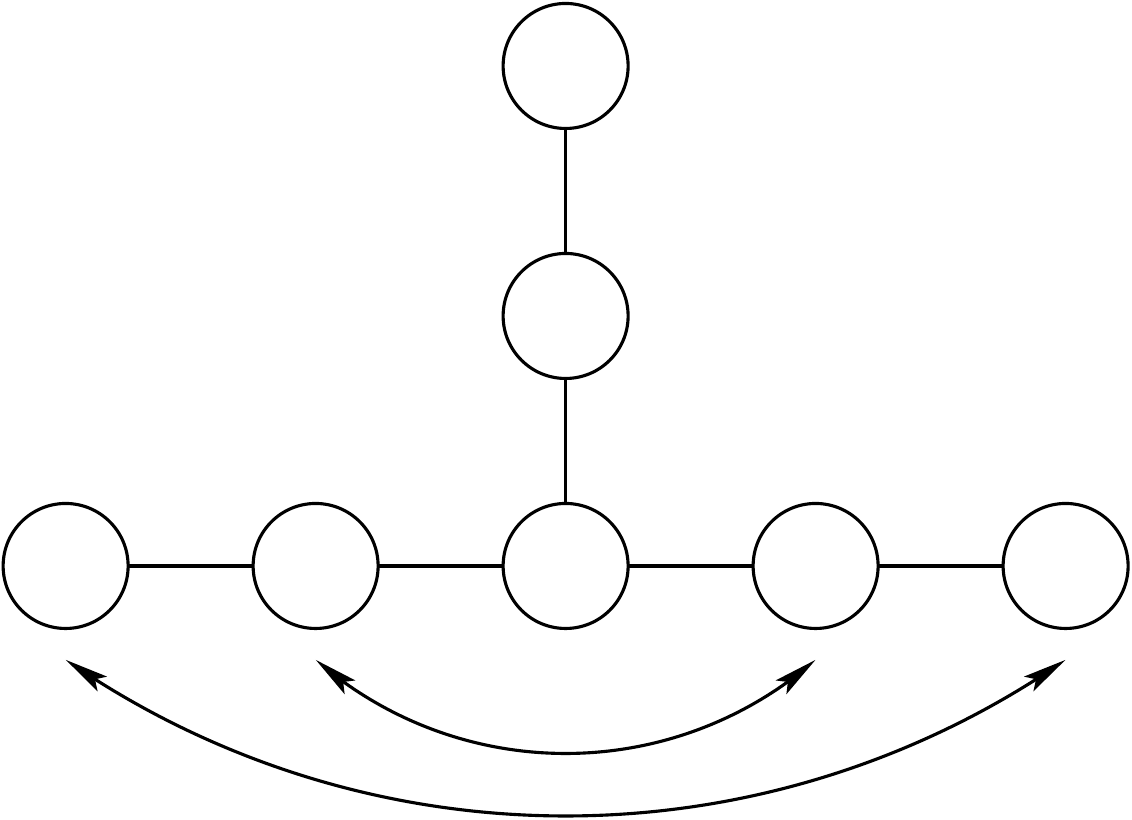}} \notag
\end{align}
\caption{Affine quiver diagrams with double-lined arrows (left) obtained by folding simply laced quivers (right). The 
double-lined arrow
is drawn from an orthosymplectic group to a unitary group.  The numbers in the circles denote the comarks or dual Kac labels and are proportional to the gauge group ranks.\label{fig:twistedADEquivers}}
\end{figure}

To treat quiver theories with orthosymplectic groups, we introduce a process we call unfolding in section \ref{ss:unfold} .  
In the large $N$ limit of the matrix model, unfolding relates
theories containing $O(N)$ and $USp(2N)$ groups to quiver theories with only $U(N)$ groups. 
For gauge theories with only $U(N)$ groups and bifundamental fields, ref.\ \cite{Gulotta:2011vp} established that
the only CS theories for which the long range forces between the eigenvalues cancel have quivers which are in one-to-one correspondence with the simply laced affine Dynkin diagrams
(see figure \ref{fig:ADEquivers}).
Given this restriction to ADE Dynkin diagrams, 
 the inverse of the unfolding procedure, which we call folding, is then 
an identification of the simply laced Dynkin diagrams under
a $\mathbb{Z}_2$ outer automorphism (see figure \ref{fig:twistedADEquivers}).

\begin{figure}[h]
\begin{align}
A^{(2)}_{2} &&& \parbox{5cm}{\scalebox{0.3}{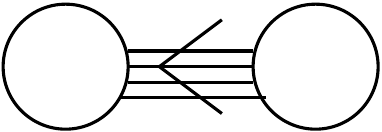}}
&&  D^{(2)}_{4} & \parbox{5cm}{\qquad\scalebox{0.3}{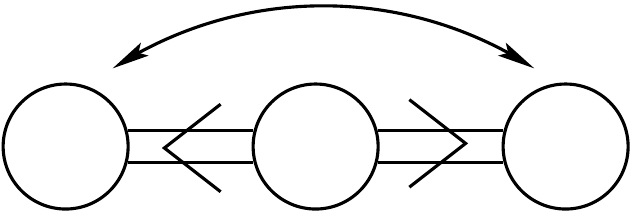}}\notag \\
A^{(2)}_{2n} &&& \parbox{5cm}{\scalebox{0.3}{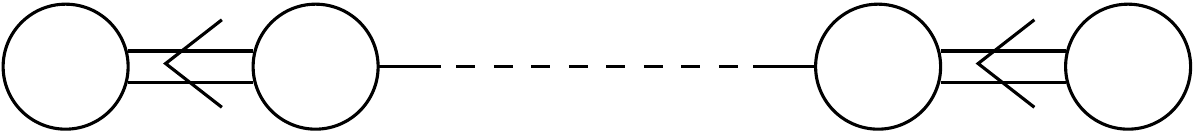}}
&&  A^{(2)}_{2n+1} & \parbox{5cm}{\qquad\scalebox{0.3}{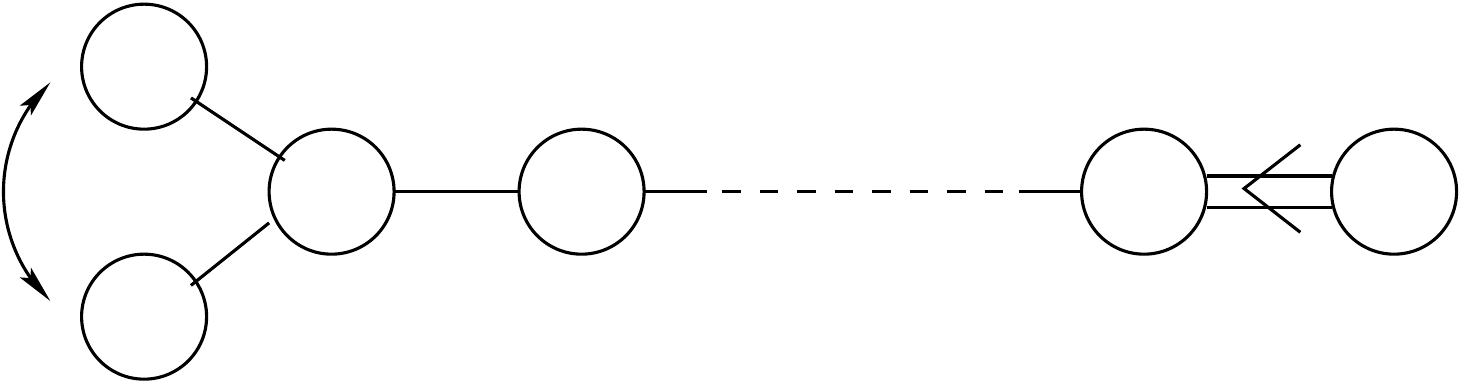}}\notag \\
D^{(3)}_{4} &&& \parbox{5cm}{\scalebox{0.3}{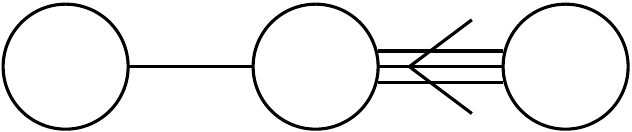}}
&& E_{6}^{(1)} & \parbox{5cm}{\scalebox{0.3}{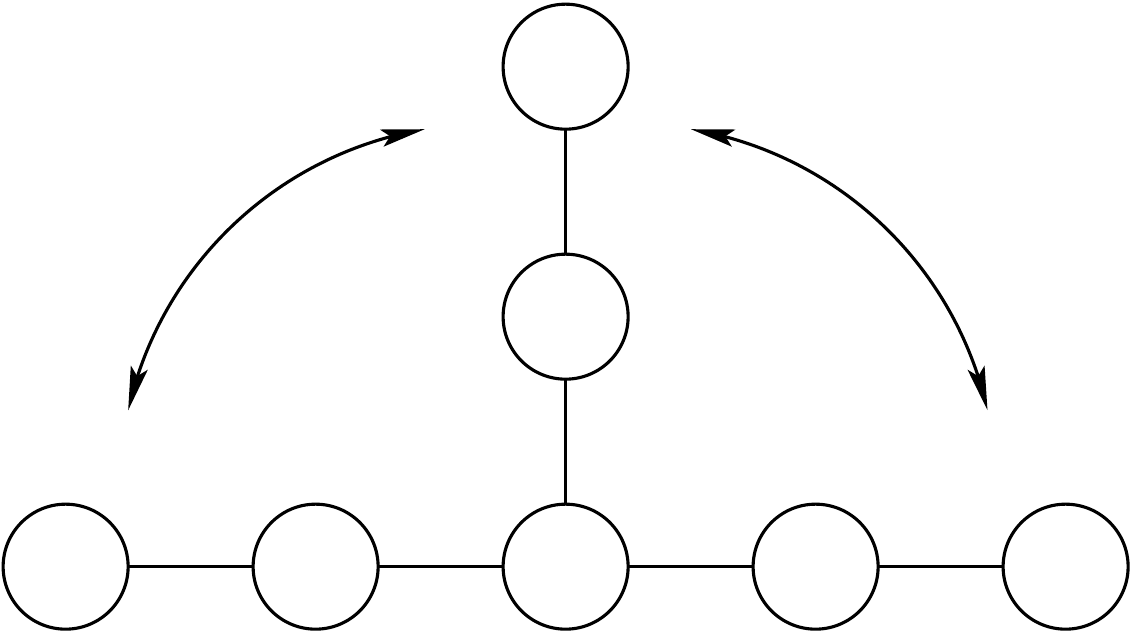}} \notag \\
G_2^{(1)} &&& \parbox{5cm}{\scalebox{0.3}{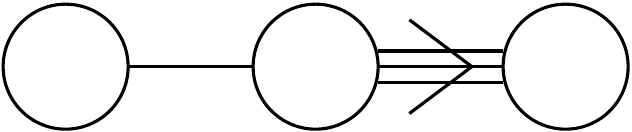}}
&&  D_{4}^{(1)} & \parbox{5cm}{\qquad\scalebox{0.3}{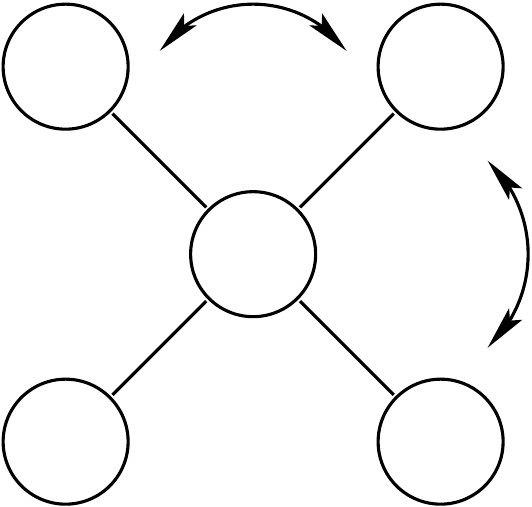}}\notag 
\end{align}
\caption{Affine Dynkin diagrams other than figure \ref{fig:twistedADEquivers}.\label{fig:twistedDynkin}}
\end{figure}

Given the restriction to ${\cal N}=3$ SUSY, classical groups, and bifundamental matter fields, we believe
that figure \ref{fig:ADEquivers} and the left hand column of figure \ref{fig:twistedADEquivers} is a complete enumeration of the CS matter theories with cancellation of long range forces between the eigenvalues
 in the large $N$ limit.  
However, we note there are larger $\mathbb{Z}_2 \times \mathbb{Z}_2$ and $\mathbb{Z}_3$ outer automorphisms of certain simply laced Dynkin diagrams for which we have no gauge theory interpretation of the corresponding folded quiver (see figure \ref{fig:twistedDynkin}).

This unfolding procedure gives a simple relationship between the free energy of the unfolded theory and the folded theory that involves some factors of two.  There is a corresponding simple relation between the volume of $Y$ and hence between the moduli spaces.
We shall examine these factors of two and how the unfolding works in a few specific examples 
in detail in section \ref{sec:examples}.

Folding and unfolding of the matrix model has a string theory representation as the addition and subtraction respectively of orientifold and orbifold planes.  In section \ref{sec:branes} we review how to construct these orthosymplectic theories from D3 branes, O3 planes, O5 planes, and orbifold planes in type IIB string theory.  Although we will not discuss it further, we mention in passing that these brane constructions can be T-dualized and uplifted to the eleven dimensional supergravity solutions discussed previously \cite{Hanany:2000fq}.

While the folding/unfolding story is the central theme of the paper, there are a few more sections of interest.
In section \ref{sec:ADEmatrix}, we relate our $U(N)$ quiver theories 
 to the $\widehat{\text{ADE}}$ quiver matrix models of refs.\ \cite{Kharchev:1992iv, Kostov:1992ie} 
 with a $\ln(x)^2$ Stieltjes-Wigert potential.  
One nice aspect of this relationship is that at the level of the matrix model it
connects the generalized Seiberg duality for the 2+1 dimensional gauge theories \cite{Aharony:1997gp,Giveon:2008zn} with Seiberg duality for a class of 3+1 dimensional gauge theories \cite{Cachazo:2001sg,Dijkgraaf:2003xk}.
The appendices collect some well known but useful facts about orthosymplectic gauge theories.
 Appendix \ref{app:reps} describes the fundamental and adjoint representations of the classical groups.
 Appendix \ref{app:CSnorm} recalls how the CS term in the action is conventionally normalized. 
 Appendix \ref{app:N3} reviews the flavor symmetries and the definition of a half hypermultiplet for 
 gauge theories with orthosymplectic groups.

%%%%%%%%%%%%%%%%%%%%%%%%%%%%%%%%%%%%
\section{Matrix models for $\CN = 3$ gauge theories}
\label{sec:matrixmodels}
%%%%%%%%%%%%%%%%%%%%%%%%%%%%%%%%%%%%
The partition function of an $\CN = 3$ SUSY Chern-Simons matter theory on an $S^3$ localizes to an eigenvalue integral \cite{Kapustin:2009kz}.
Let us review how to construct this integral for a gauge group that is a direct product of simple compact Lie groups, $G = \otimes_a G_a$.
The eigenvalues in question are the eigenvalues of the auxiliary scalars $\sigma_a$ in the vector multiplet (see appendix \ref{app:N3} for our conventions).  Let us denote the eigenvalues of $\sigma_a$ by
$\mu_{a,i}$, $i = 1, \ldots, N_a$.

The vector multiplets and the matter fields contribute separately to the partition function,
\begin{align}
\label{ZS3}
Z &=  \, \int \left( \prod_{a,l} \frac{\d \mu_{a,l}}{\operatorname{rank}(\calW_a)}  \right) \left( \prod_a L_V(G_a, k_a, \mu_a) \right) 
\left( \prod_I L_M (\calR_I, \mu)\right)   \\
 &=  \, \int \left( \prod_{a,l} \frac{ \d \mu_{a,l}}{\operatorname{rank}(\calW_a)}  \right) \exp \left(-F(\{\mu_{a,l} \}) \right) \ ,
\end{align}
where $k_a$ is the CS level and $\calW_a$ the Weyl group associated to $G_a$ \cite{Kapustin:2009kz}.
The $\operatorname{rank}(\calW_a)$ normalization factor  will turn out to be subleading in our large $N$ expansion.  We give the ranks of the Weyl groups along with some other representation theory
data for the classical groups in appendix \ref{app:reps}.
We denote by $L_V(G_a, k_a)$ the vector multiplet contribution from gauge group $G_a$, and by $L_M(\calR_I)$ the hypermultiplet contribution from the representations $\calR_I$ and $\calR_I^*$.

More precisely, the vector multiplet contribution is 
\begin{align}
	L_V (G, k, \mu) =  e^{i\pi k \mu^2} \prod_{\alpha >0}(2\sinh [\pi \alpha \cdot \mu])^2 \ ,
\end{align}
where $\mu$ is a weight vector, and $\alpha$ is a positive root.
We normalize the $\alpha$ such that the longest root has length squared equal to two.  We expand $\mu_a = (\mu_{a,1}, \mu_{a,2}, \ldots, \mu_{a,N})/c_a$ in terms of an orthonormal basis on the weight lattice of $G_a$ and choose the normalization constant $c_a$ to ensure that some arguments of the hyperbolic sine function have the form $\pi(\mu_{a,l} - \mu_{a,m})$.   
(Note the rescaling will change the normalization of the measure factor, but we ignore this rescaling because it will be subleading in $N$.)
For
$U$, $O$ and $USp$ groups we obtain
\be
{\renewcommand{\arraystretch}{1.8}
\begin{array}{c|c}
G_a & L_V \\
\hline
U(N)_k &
e^{i\pi k \sum_{m=1}^N \mu_m^2} \prod_{l<m}(2\sinh [\pi (\mu_l - \mu_m)])^2
\\
\hline
O(2N)_k &
e^{i\pi k \sum_{m=1}^N \mu_m^2} \prod_{l<m}(4\sinh [\pi (\mu_l + \mu_m)]
		\sinh [\pi (\mu_l - \mu_m)])^2
\\
\hline
O(2N+1)_k &
e^{i\pi k \sum_{m=1}^N \mu_m^2} \prod_{l<m}(4\sinh [\pi (\mu_l + \mu_m)]
		\sinh [\pi (\mu_l - \mu_m)])^2
		 \prod_m (2\sinh[\pi \mu_m])^2
\\
\hline
USp(2N)_k & e^{i 2\pi k \sum_{m=1}^N \mu_m^2} \prod_{l<m}(4\sinh [\pi (\mu_l + \mu_m)]
			\sinh [\pi (\mu_l - \mu_m)])^2
			 \prod_m (2\sinh[2\pi \mu_m])^2\\
\hline
\end{array}
}
\ee
For $U$ and $USp$ groups, we anticipate that $k$ is an integer, while for $O$ groups, $k$ must be an even integer (see appendix \ref{app:CSnorm}).

The contribution from a hypermultiplet is
\begin{align}
	L_M (\calR, \mu) = \prod_{\rho \in \calR} \frac{1}{2\cosh [\pi \rho \cdot \mu]} \ ,
\end{align}
where $\rho$ is a weight vector in the representation $\calR$ (not the Weyl vector).
In this paper, we are interested in bifundamental representations of the classical groups $U$, $USp$, and $O$.  
Three different types of $L_M$ arise for us:
\be
{\renewcommand{\arraystretch}{1.5}
\begin{array}{c|c}
\calR & L_M \\
\hline
U(N_a) \times U(N_b) &
\left( \prod_{l,m} 2 \cosh [ \pi (\mu_{a,l} - \mu_{b,m} )] \right)^{-1} 
\\
\hline
U(N_a) \times O(2N_b), 
U(N_a) \times USp(2N_b), &
\left( \prod_{l,m} 
4 \cosh [ \pi (\mu_{a,l} - \mu_{b,m}) ] \cosh [ \pi (\mu_{a,l} + \mu_{b,m}) ] \right)^{-1} 
\\
O(2N_a) \times USp(2N_b)  
&
\\
\hline
U(N_a) \times O(2N_b+1), USp(2N_a) \times O(2N_b+1)
 &
\left( \prod_{l,m} 
4 \cosh [ \pi (\mu_{a,l} - \mu_{b,m}) ] \cosh [ \pi (\mu_{a,l} + \mu_{b,m}) ] \right)^{-1}
\\
&
\times \left( \prod_l 2 \cosh [ \pi \mu_{a,l} ] \right)^{-1}
\\
\hline
\end{array}
}
\label{Lm}
\ee
The matter fields involving a $U$ group are hypermultiplets under the supersymmetry algebra of the CS theory.  For $O \times USp$ bifundamentals, the rules giving the contribution of a hypermultiplet would naively give the square of the result in the right hand column because the fundamental weights appear in pairs $\pm {\bf e}_i$ where ${\bf e}_i \in {\mathbb R}^N$ are orthonormal basis vectors of the weight lattice.  In these cases, having taken a square root, what we have listed in the right hand column is the contribution of a half hypermultiplet.  Half hypermultiplets exist for $O \times USp$ bifundamentals because the product representation is pseudoreal (see appendix \ref{app:N3}). 
Recall that $N_f$ hypermultiplets for an $O$ group have an enhanced 
$USp(2N_f)$ global symmetry for $\calN = 3$ SUSY theories. 
Similarly, $N_f$ half hypermultiplets under a $USp$ group transform under an $O(N_f)$ global symmetry.
(For more details, see appendix \ref{app:N3} or  \cite{Argyres:1995fw}.)
Conveniently 
for $O(N) \times USp(2M)$ bifundamentals, these two global symmetries are completely gauged.

Note that we have not given the results for $O \times O$ or $USp \times USp$ bifundamentals;
$L_M$ for such a representation is roughly the square of $L_M$ for a $O \times USp$ half hypermultiplet. 
Because the product representation is real, these $O \times O$ and $USp \times USp$ 
bifundamentals cannot be half hypermultiplets. 
The fact that $O(N) \subset USp(2N) \subset O(4N)$ allows one to gauge the appropriate subgroup of the 
global flavor symmetry group for these bifundamentals.  
It will turn 
out the only quivers that contain such representations and that 
satisfy the long range force cancellation condition we discuss below are
of $A_1^{(1)}$ type, i.e.\ $O(N)^2$ or $USp(2N)^2$ with a full bifundamental hypermultiplet between the groups.  We will see in section \ref{sec:branes} how this type of gauge theory can arise in 
a type IIB brane construction of the $C_1^{(1)}$ quiver.

We henceforth will consider arbitrary theories consisting of products of $U$, $USp$, and $O$ gauge groups and bifundamental fields of the type listed in (\ref{Lm}).  Thus, our field theory can be specified by a quiver, or equivalently a collection of nodes $V$ and edges $E$.  Each node corresponds to a gauge group factor $G_a$.  
Each edge corresponds to a 
pair of bifundamental representations $\calR$ and $\calR^*$.  When $\calR \neq \calR^*$, an edge means a full hypermultiplet.  When $\calR = \calR^*$, an edge means a half hypermultiplet (except in the $O \times O$ and $USp \times USp$ cases).

%%%%%%%%%%%%%%%%%%%%%%%%%%%%%%%%%%%%
\subsection{Unfolding trick}\label{ss:unfold}
%%%%%%%%%%%%%%%%%%%%%%%%%%%%%%%%%%%%
We present a trick we call unfolding which, in the large $N$ limit, reduces a matrix model involving  
$O/USp$ groups to a matrix model involving only $U$ groups.
The unfolding produces a quiver theory with a ${\mathbb Z}_2$ symmetry.
An $O(N)$ or $USp(N)$ group is converted into a $U(N)$ group. (For $USp$, $N$ should be even.)  Each $U(N)$ group is converted into a pair of $U(N)$ groups.  A half hypermultiplet between an $O$ and $USp$ group is lifted to a full hypermultiplet between two $U$ groups.  A hypermultiplet between an $O/USp$ and a $U$ group is lifted to two hypermultiplets between the $U$ uplift of the $O/USp$ and two $U$ groups.  A hypermultiplet between two $U$ groups is lifted to a pair of hypermultiplets.

Let us show how this unfolding works in detail.  
For the $U(N)$ vector multiplets, we duplicate the eigenvalues, introducing $\lambda = \mu$ and $\lambda' = \mu$:
\be
L_V(U(N), k, \mu) = L_V(U(N), k, \lambda)^{1/2}  L_V(U(N), k, \lambda')^{1/2} \ .
\label{vectorunfold}
\ee
For $O(2N)$, $O(2N+1)$, and $USp(2N)$ vector multiplets, we define a set of eigenvalues that is twice as large, $\lambda_l = \mu_l$ and $\lambda_{l+N} = - \mu_l$.  For $O(2N+1)$, the unfolded $U(2N+1)$ has an additional zero eigenvalue $\lambda_{2N+1}=0$.  Up to an overall $-1$ that we will not keep track of,
\begin{align}
L_V(O(N), k, \mu) &= (-1)^\# L_V(U(N), k, \lambda)^{1/2} \left( \prod_{m} 2 \sinh[2 \pi \lambda_m] \right)^{-1/2} \ , \label{vONunfold}\\
L_V(USp(2N), k, \mu) &= (-1)^\# L_V(U(2N), 2k, \lambda)^{1/2} \left( \prod_{m} 2 \sinh[2 \pi \lambda_m] \right)^{1/2} \ .  \label{vUSpNunfold}
\end{align}
With the same conventions for the eigenvalues, we can also write down how the hypermultiplets unfold
\be
{\renewcommand{\arraystretch}{1.5}
\begin{array}{c|c}
\calR & L_M \\
\hline
U(N_a) \times U(N_b) &
\left( \prod_{l,m} 4 \cosh [ \pi (\lambda_{a,l} - \lambda_{b,m} )]
\cosh [ \pi (\lambda'_{a,l} - \lambda'_{b,m} )] \right)^{-1/2} 
\\
\hline
U(N_a) \times O(N_b), 
U(N_a) \times USp(2N_b), &
\left( \prod_{l,m} 
4 \cosh [ \pi (\lambda_{a,l} - \lambda_{b,m}) ] \cosh [ \pi (\lambda'_{a,l} - \lambda_{b,m}) ] \right)^{-1/2} 
\\
\hline
O(N_a) \times USp(2N_b)  
&
\left( \prod_{l,m} 
2 \cosh [ \pi (\lambda_{a,l} - \lambda_{b,m}) ] \right)^{-1/2} 
\\
\hline
\end{array}
}
\label{Lmunfolded}
\ee

We would like to argue that in a large $N$ limit, the difference between an unfolded matrix model and a matrix model constructed out of a product $\otimes_a U(N_a)$ is subleading in $N$.  
More specifically, taking advantage of a large $N$ limit to perform a saddlepoint integration, we will argue that the saddlepoint eigenvalue distributions are identical.  
Recall the distributions are determined by the saddlepoint equations
$\partial F / \partial \mu_{a,m} = 0$ before unfolding and $\partial F / \partial \lambda_{a,m} = 0$ after unfolding.

One obvious difference between the unfolded model and a $\otimes_a U(N_a)$ model is the eigenvalue integration measure.  In the unfolded model, we have constraints on the eigenvalues, 
while in an $\otimes_a U(N_a)$ model, there are no such constraints. 
In the cases we have examined numerically,
the constraints are respected by the eigenvalue distribution of the $\otimes_a U(N_a)$ model.
From an analytic perspective, 
we can prove the constraints are obeyed modulo a uniqueness assumption.
The constraint we placed on an unfolded $O/USp$ 
node was that the eigenvalue distribution of the $U(2N)$ group be symmetric about the origin. 
Indeed,  if the set
$\{ \lambda_{a,i} \}$ satisfies the saddlepoint equations, then so does 
the set $\{ - \lambda_{a,i} \}$.  
Thus if there is only one solution to the saddlepoint equations, then the constraint will be obeyed.
In unfolding a $U(N)$ group to $U(N)^2$, we created two sets of identical eigenvalues.  
Because of the ${\mathbb Z}_2$ symmetry of the unfolded quiver, the saddlepoint equations for the eigenvalues in the two $U(N)$ groups will be the same and the eigenvalue distributions will be identical, again provided there is a unique solution.

A further obvious difference are the square roots that appear in the unfolded versions of the
 $L_V$ and $L_M$ factors.  In a saddlepoint analysis, these square roots assemble to become an overall 1/2 multiplying the free energy $F$.  We need to keep track of this 1/2, but it will not change the form of the saddlepoint eigenvalue distribution.  

A final troubling difference are the 
factors of $\sinh[ 2 \pi \lambda_{a,l}]$.
At worst, in our large $N$ limit 
they contribute in the same way as flavor fields in the fundamental representation, but
it turns out that these factors will largely cancel at leading nontrivial order in $N$.  To see this cancellation,  we first need to review the large $N$ analysis of the 
saddlepoint equations $\partial F / \partial \lambda_{a,m} = 0$ performed in \cite{Gulotta:2011vp}. 

The key ingredient in being able to find a solution to the saddle point equations and a corresponding eigenvalue distribution is the cancellation in the long range forces between the eigenvalues.   
Given a gauge group $\otimes_a U(N_a)$, 
let $N_a = n_a N$ where the $n_a$ are relatively prime integers and $N \gg 1$.  
We take a continuum large $N$ limit by assuming that the eigenvalues lie along the curves
\be
\lambda_{a,I}(x) = N^{\alpha} x + i y_{a,I}(x) \ ,
\label{eigenvalueansatz}
\ee
$I = 1, 2, \ldots, n_a$, 
characterized by an eigenvalue density $\rho(x)$ such that  $\int \rho(x) dx = 1$.  

We examine the saddlepoint equation for an eigenvalue in the $\lambda_{a,I}$ curve.
At leading order in $N$, we can replace the $\tanh$ and $\coth$ functions that appear in the saddle point equations with sign functions.  
Because of the double sum, only the $\tanh$ and $\coth$ functions that involve differences
$\lambda_{a,l} - \lambda_{b,m}$ of eigenvalues will appear at this leading order in $N$.  
The contributions from the $\sinh [ 2 \pi \lambda_{a,l}]$ factors 
are suppressed by one additional power of $N$.
In the continuum limit, the leading order in $N$ term in the saddle point equation for an 
eigenvalue in the $\lambda_{a,I}$ curve is
\be
\left( 2 n_a  - \sum_{b | (a,b) \in E} n_b  \right) N
\int dx' \sgn(x-x') \rho(x') \ ,
\ee
We call such a force on the eigenvalue long range because there is a contribution from non-neighboring eigenvalues.  
Thus we conclude that
\be
2 n_a  = \sum_{b | (a,b) \in E} n_b  \ .
\label{nf2nc}
\ee
As noted in \cite{Gulotta:2011vp}, this condition implies 
that the quivers for these $\otimes_a U(N_a)$ theories must 
be affine ADE Dynkin diagrams (see figure \ref{fig:ADEquivers}).  
In more detail, 
note that the condition \eqref{nf2nc} can be written as $\sum_b\hat A_{ab} n_b = 0$, where $\hat A$ is an affine Cartan matrix of ADE type:
$\hat A$ is symmetric with diagonal entries $\hat A_{aa}=2$ and non-zero off-diagonal entries $\hat A_{ab}=-1$.
Affine Cartan matrices $\hat A$ have a one dimensional kernel, and 
it turns out that the ranks of the gauge groups $n_b$ are the comarks or dual 
Kac labels \cite{Kac:1990gs,DiFrancesco:1997nk} shown in figure \ref{fig:ADEquivers}.
A corresponding force cancellation condition can be derived directly for general quiver theories with both unitary and orthosymplectic groups.  
The condition is still $\sum_b \hat A_{ab} n_b= 0$ but now $\hat A_{ab} = -2$ if $a$ is $O/USp$ type and $b$ is $U$ type.  Otherwise the off-diagonal entries are $0$ or $-1$ as before.

Before returning to the main thread of our argument, we make two remarks.
Although for simplicity we consider here only the case of $S^3$, a similar analysis for $\CN =2$ gauge theories on a squashed three-sphere \cite{Hama:2011ea,Imamura:2011wg}
leads to the same constraint \eqref{nf2nc} in a large $N$ limit.
In the introduction, we mentioned a secondary condition necessary for the existence of an $AdS_4 \times Y$ dual gravity description.  If the CS levels of the unfolded quiver are $k_a$, 
this condition  is
\cite{Herzog:2010hf, Jafferis:2011zi}
\be
\sum_a n_a k_a = 0 \ .
\ee

Returning now to the extra factors of $\sinh [ 2 \pi \lambda_{a,l}]$ in the unfolded matrix model, note first that these factors occur only for $O$ and $USp$ type groups.
An $O(N_a)$ group contributes a $\sinh [2 \pi \lambda_{a,l}]$ to the denominator of the partition function which at large $N$ has the same effect as adding a fundamental flavor hypermultiplet.  
A $USp(2N_b)$ group contributes a $\sinh [2 \pi \lambda_{b,l}]$ to the numerator, which would exactly cancel the contribution from a flavor hypermultiplet at leading order in $N$.  
More specifically, if the original theory had an $(\otimes_a O (N_a)) (\otimes_b USp(N_b))$ gauge group factor, then the free energy would have a term proportional to 
\be
\left(\sum_a N_a- \sum_b N_b \right) \sgn(x) \ ,
\ee
 where $N_{\rm eff} = (\sum_a N_a- \sum_b N_b)/N$ is like an effective number of flavors.  

If we want the unfolded model to match exactly a $\otimes_a U(N_a)$  model (with no extra fundamental fields) at leading order in $N$, then we must always take $N_{\rm eff} \leq 0$.  In this way, we can make up the discrepancy in the free energy by adding a few extra fundamental hypermultiplets to the original orthosymplectic theory to cancel the effect of the $\sinh [2 \pi \lambda_{a,l}]$ in the numerator.  Interestingly, this type of restriction on the 
number of $USp$ and $O$ groups was discussed in ref.\ \cite{Uranga:1998uj} from the point of view of brane constructions and tadpole cancellation.  We will come back to this point at the end of section \ref{sec:branes} which discusses brane constructions for some of these theories.

Let us see for what types of quivers we need to worry about a nonzero $N_{\rm eff}$.  
If we take an $A_n^{(1)}$ ($n>1$) quiver where all of the $U(N)$ groups have been replaced with $O$ and $USp$ groups, then in order to avoid having a half hypermultiplet in $O \times O$ or $USp \times USp$ representations, we need to take $n$ odd.  But in this case, the number of $O$ and $USp$ groups will be equal and $N_{\rm eff}=0$.

For the $D_n^{(1)}$ quivers where all of the $U(N)$ groups are replaced with $O$ and $USp$ groups, we find two cases.  For $n$ odd, $N_{\rm eff} = 0$ and the unfolded model 
matches the $\otimes_a U(N_a)$ model.  For $n$ even, $N_{\rm eff} = \pm 2$.  In the case where we have a couple of extra $USp$ gauge groups, we find a minus sign which we could cancel by adding a couple of fundamental hypermultiplets.  Similarly, for $E_6^{(1)}$, we find $N_{\rm eff} = 0$ while for
$E_7^{(1)}$ and $E_8^{(1)}$, $N_{\rm eff} = \pm 2$.  
Continuing down the list, we see that $B_n^{(1)}$, $C_n^{(1)}$ and $D_{n+1}^{(2)}$ theories have $N_{\rm eff} = 0$, $\pm 2$, while $A_{2n-1}^{(2)}$, $E_6^{(2)}$, and $F_4^{(1)}$ theories have $N_{\rm eff} = \pm 2$.  As we will discuss in section \ref{sec:branes}, the fact that the discrepancy $N_{\rm eff}$ is either 0, $-2$ or 2 is related to the charge of an O5 plane in a brane construction of the $A$, $B$, $C$, and $D$ type gauge theories.

%%%%%%%%%%%%%%%%%%%%%%%%%%%%%%%%%%%%
\subsection{Unfolding and $\BZ_2$ automorphism of affine Dynkin diagrams}\label{ss:unfoldDynkin}
%%%%%%%%%%%%%%%%%%%%%%%%%%%%%%%%%%%%

Restricting to affine ADE Dynkin diagrams, we discover that the procedure we have called unfolding has a long history.  (See for example \cite{Fuchs:1995zr} for an application to 2d conformal field theory.)  
The inverse of this unfolding procedure, let us call folding,
takes advantage of a 
 ${\mathbb Z}_2$ outer automorphism of the Dynkin diagram.  The simplest type of folding operation is one in which every $U$ node in an affine ADE Dynkin diagram is replaced by an $O/USp$ type node. 
More complicated folding procedures are also allowed.   
Let us start with the third row of figure \ref{fig:twistedADEquivers} and an $A_{2n-1}^{(1)}$ quiver.  
We pairwise identify $2n-2$ of the $U(N)^{2n}$ groups, folding them to obtain $n-1$ $U(N)$ groups.  In this identification, 
we need to make sure that two $U(N)$ groups we combine to form a single $U(N)$ group have the same CS level.
The remaining two $U(N)$ groups at the ends are folded to obtain $O/USp(N)$ groups. 
In the folding procedure, the pairwise identification of the edges of the $A_{2n-1}^{(1)}$ quiver is a corresponding identification of the bifundamental hypermultiplets.

The other examples in figure \ref{fig:twistedADEquivers} fold in a similar way.
For the $A_{2n-1}^{(2)}$ quiver, the node at the right edge with orthosymplectic
group comes from a node in the middle of the $D_{2n}^{(1)}$ quiver.
The remaining nodes in the $D_{2n}^{(1)}$ quiver are pairwise identified and fold down to unitary groups
in $A_{2n-1}^{(2)}$.
For $B_n^{(1)}$ and $D_{n+1}^{(2)}$, the nodes at the edges with unitary group 
descend from pairs of unitary nodes from the forked ends of 
$D_{n+1}^{(1)}$ and $D_n^{(1)}$ respectively.
The remaining unitary groups in $D_{n+1}^{(1)}$ and $D_n^{(1)}$ fold to orthosymplectic groups.
For $E_6^{(2)}$ and $F_4^{(1)}$, there is a short chain of unitary groups which descends from a pair
of chains in $E_7^{(1)}$ and $E_6^{(1)}$ respectively.

We believe that figures \ref{fig:ADEquivers} and \ref{fig:twistedADEquivers} give a 
complete list of $\calN=3$ SUSY CS bifundamental matter theories where the long range forces cancel in the corresponding matrix model.  In figure  \ref{fig:ADEquivers}, the nodes can be interpreted either as $U(N)$ groups or as alternating $O/USp$ groups.  (Note that for the $A_n^{(1)}$ quivers ($n>1$), $n$ must be odd in order to avoid having an $O \times O$ or $USp \times USp$ half hypermultiplet.)
All our non-simply laced examples come in pairs: $A_{2n-1}^{(1)}$ and $B_n^{(1)}$, $C_n^{(1)}$ and $D_{n+1}^{(2)}$, and $E_6^{(2)}$ and $F_4^{(1)}$.  Note that the Cartan matrix for one member of the pair is the transpose of the Cartan matrix for the other.  The simply laced examples, in contrast, have symmetric Cartan matrices.

Certain simply laced Dynkin diagrams have an outer automorphism group 
that is larger than $\mathbb{Z}_2$.  For example, $E_6^{(1)}$ and $D_4^{(1)}$ can 
both be folded using a $\mathbb{Z}_3$, as shown in figure \ref{fig:twistedDynkin}.  
It would be interesting to understand if the resulting folded Dynkin diagrams have a quiver gauge theory interpretation.

Before ending this section, let us consider in more detail folding an $A_{2n-1}^{(1)}$ $U(N)^{2n}$ quiver which according to the prescription outlined thus far would yield an 
$[O(N) \otimes USp(N)]^n$ quiver.  The appearance of a $USp(N)$ would naively seem to suggest that $N$ must be even.  However, we can modify slightly the unfolding above to allow for an odd $N$.  In particular, we can unfold a $USp(2N_a)$ quiver to a $U(2N_a+1)$ quiver instead of a $U(2N_a)$ quiver by allowing for a $\lambda_{2N_a+1} = 0$ eigenvalue.  This extra eigenvalue will introduce an extra factor of
$\prod_m \sinh [\pi \lambda_{a,m}]^{-1/2}$ in the vector multiplet and extra factors of 
$\prod_{b | (a,b) \in E} \prod_m \cosh [\pi \lambda_{b,m}]^{1/2}$ in the hypermultiplets.  Because of the condition (\ref{nf2nc}), these factors will cancel out at leading nontrivial order in $N$.  
In this way, we can fold an $A_{2n-1}^{(1)}$ $U(2N+1)^{2n}$ quiver to an 
$[O(2N+1) \otimes USp'(2N)]^n$ quiver where we have marked the $USp$ group with a $'$ for reasons that will become clearer after we discuss the brane constructions.

%%%%%%%%%%%%%%%%%%%%%%%%%%%%%%%%
\subsection{More on unfolding}\label{ss:unfoldDn}
%%%%%%%%%%%%%%%%%%%%%%%%%%%%%%%%
The type of unfolding prescription investigated above can also be applied to a certain class of $D_n^{(1)}$ 
$U(N)^4 \times U(2N)^{n-3}$ 
quivers where the CS levels are restricted, converting the $D_n^{(1)}$ quiver into an $A_{2n-5}^{(1)}$ quiver. 
The restriction is that 
the two $U(N)$ gauge groups at a forked end of the $D_n^{(1)}$ quiver must have equal CS levels.  
The unfolding procedure will replace these two $U(N)_k$ groups with a $U(2N)_{2k}$ group.  In more detail, let the eigenvalues of the two $U(N)$ groups be $\mu_l$ and $\mu'_l$.  The eigenvalues of the $U(2N)$ 
group will then be $\lambda_l = \mu_l$ and $\lambda_{l+N} = \mu'_l$.    With these assignments, we can rewrite the contribution of the $U(N)^2$ vector multiplets as
\be
L_V(U(N), k, \mu) L_V(U(N), k, \mu') = L_V(U(2N), 2k, \lambda)^{1/2} \ .
\label{vDnunfold}
\ee
The vector multiplets for the $U(2N)^{n-3}$ groups (and the hypermultiplets between them) 
unfold just as they did above in  (\ref{vectorunfold}) and (\ref{Lmunfolded}).
It remains to specify how the hypermultiplets at the forked ends of the $D_n^{(1)}$ quiver unfold.
Let us label the eigenvalues of the $U(2N)$ group neighboring the $U(N)^2$ as $\nu_m$ and the unfolded eigenvalues of the $U(2N)^2$ group as $\kappa_l$ and $\kappa'_l$.
we can rewrite the contribution from the hypermultiplets:
\begin{eqnarray}
\lefteqn{L_M(U(N) \times U(2N), \mu, \nu) L_M(U(N) \times U(2N), \mu', \nu)
= }
\nonumber
\\
&& 
L_M(U(2N) \times U(2N), \lambda, \kappa)^{1/2} L_M(U(2N) \times U(2N), \lambda, \kappa')^{1/2} \ .
\end{eqnarray}

This unfolding procedure suggests that the free energy of such a $D_n^{(1)}$ quiver theory (with the corresponding restriction on the CS levels) will be half that of the corresponding unfolded $U(2N)^{2n-4}$ $A_{2n-5}^{(1)}$ quiver theory.  Not every $A_{2n-5}^{(1)}$ quiver will fold down to a $D_n^{(1)}$ quiver.  One has to make sure first that the CS levels of the pairs of $U(2N)$ groups that get folded together are equal.

We could now combine this additional kind of unfolding with the unfolding procedures described above to find more relations between the Dynkin diagrams.  For example, we can unfold the $B_n^{(1)}$ quiver to an $A_{2n-3}^{(1)}$ quiver, assuming that the CS levels of the $U(N)^2$ groups at the forked end of the $B_n^{(1)}$ quiver are equal.  Given the restriction on the CS levels, however, this new type of unfolding seems to be less general.

We make one more comment in passing.  The folding and unfolding operations we have carried out so far involve a ``crease'' at the the nodes of the quiver.  One can ask what happens if one puts the crease on an edge.  The answer is that folding on an edge introduces a new kind of matter field in a two index representation of the neighboring gauge group \cite{Douglas:1996sw}.  As we restricted initially to quivers with only bifundamental matter fields, we did not need to consider unfolding involving these more exotic representations.

%%%%%%%%%%%%%%%%%%%%%%%%%%%%%%%
\section{Brane constructions}\label{sec:branes}
%%%%%%%%%%%%%%%%%%%%%%%%%%%%%%%
The $A$, $B$, $C$, and $D$ type quiver theories that we discussed above can for the most part be constructed from D-branes, orientifold planes, and orbifold planes in IIB string theory.  
In this context, the folding procedures we outlined above amount to the addition of orientifold or orbifold planes to a brane configuration with a $\mathbb{Z}_2$ symmetry.  
Below, we review the rules behind these constructions but do not explain where the rules come from.  
For derivations, we refer the reader to 
\cite{Gimon:1996rq,Uranga:1998uj,Kapustin:1998fa,Feng:2000eq,Gaiotto:2008ak,Hanany:1999sj,Hanany:1997gh}.

\begin{figure}
\begin{align}
A_3^{(1)} &&& \parbox{5cm}{\scalebox{0.5}{\qquad\quad 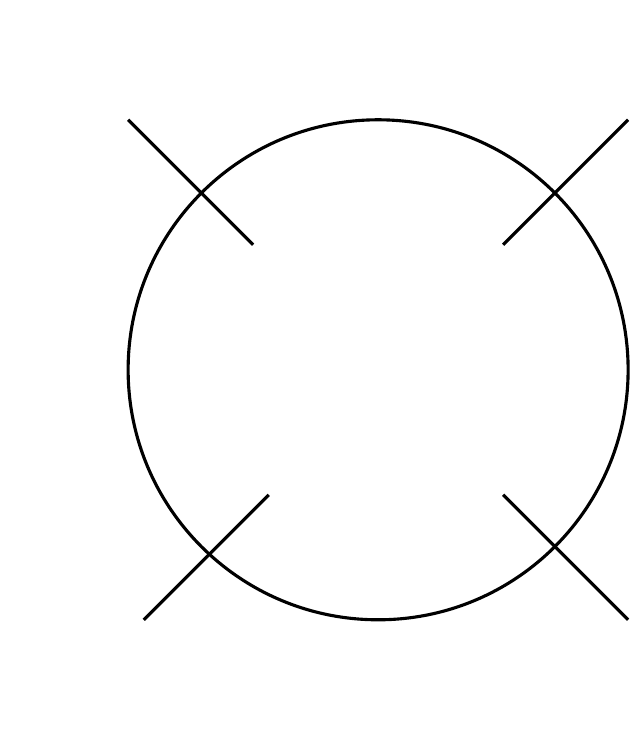}} 
&&  &\parbox{5cm}{\scalebox{0.5}{\qquad\quad 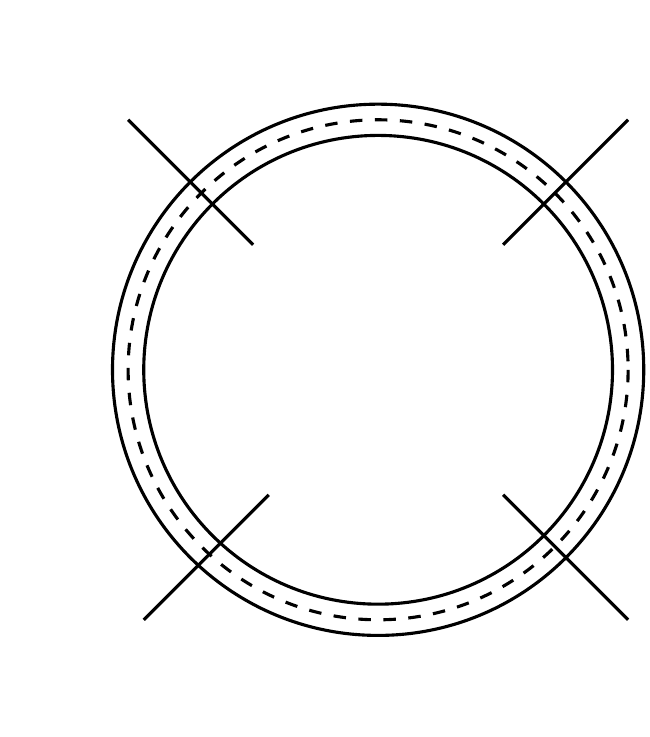}} \notag\\
A_{5}^{(2)}&&& \parbox{5cm}{\scalebox{0.5}{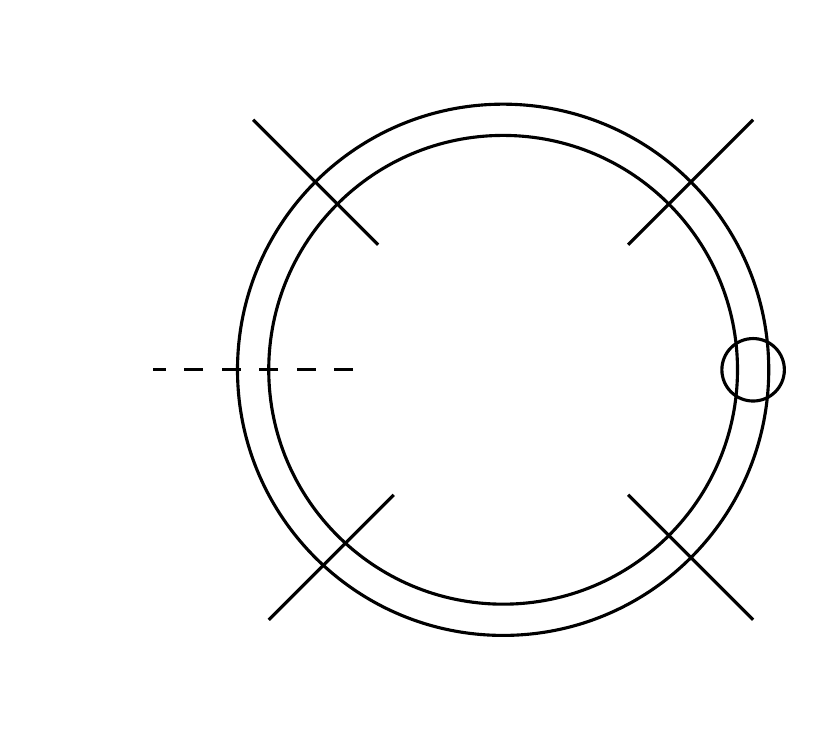}} 
&&  \qquad B_3^{(1)} & \parbox{5cm}{\scalebox{0.5}{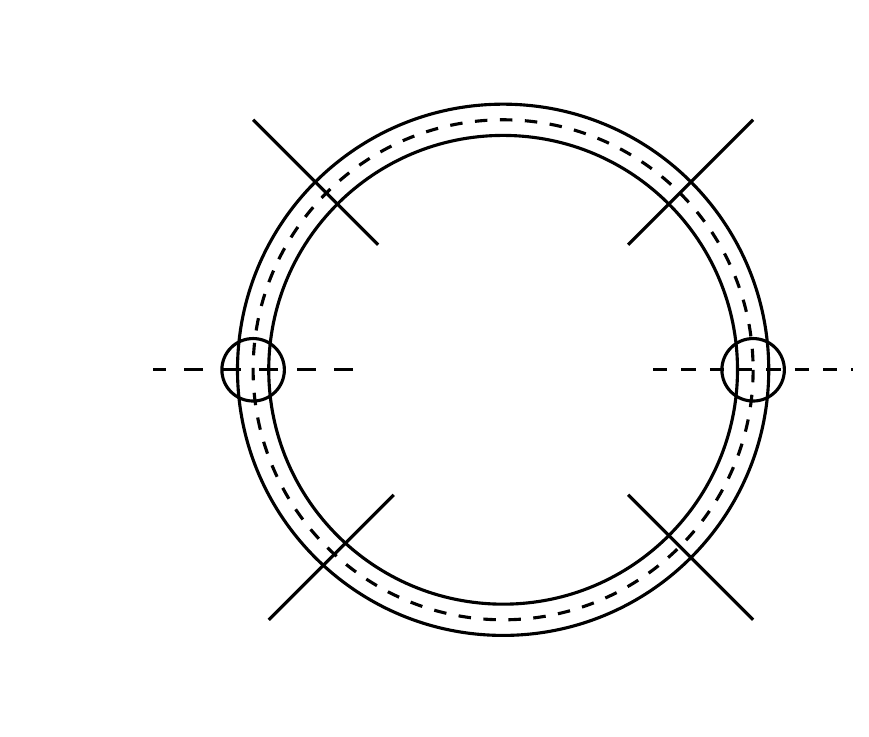}}  
\notag\\
C_2^{(1)} &&& \parbox{5cm}{\scalebox{0.5}{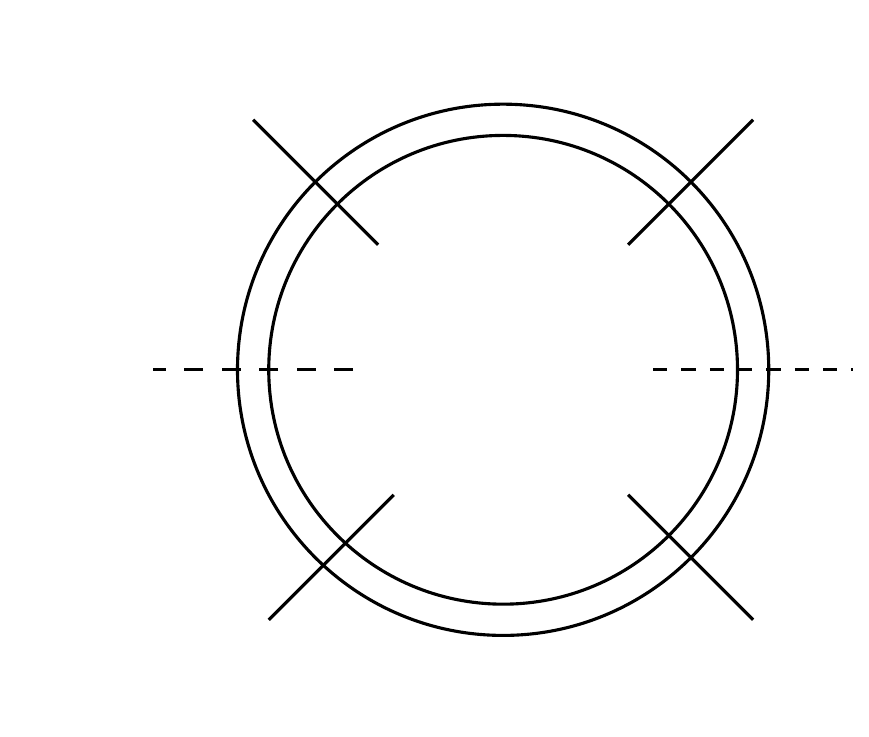}} 
&&  \qquad D_{5}^{(2)} & \parbox{5cm}{\scalebox{0.5}{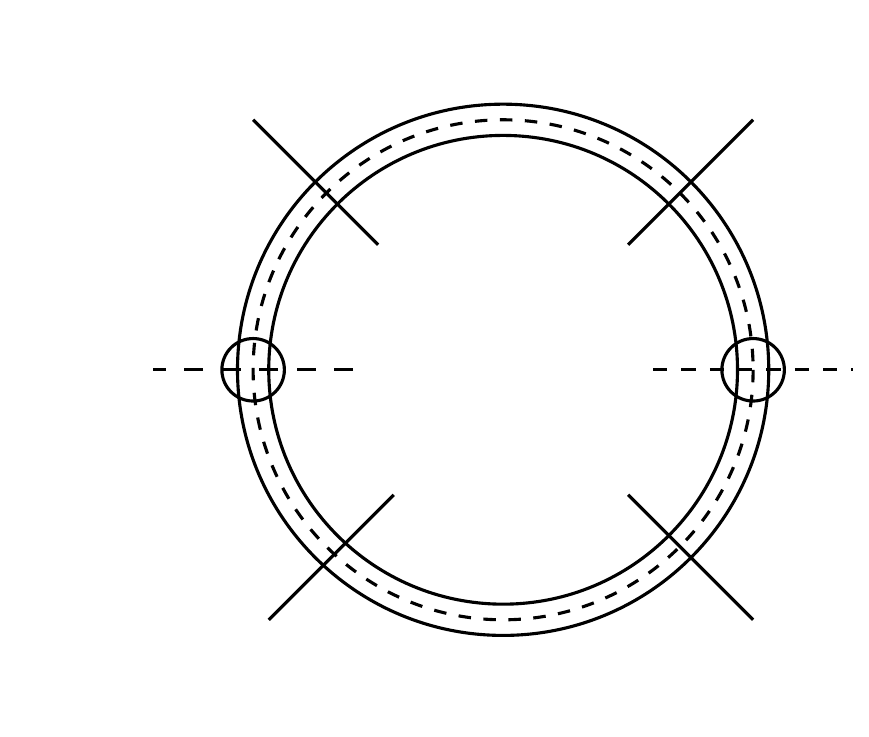}} \notag\\
D_4 ^{(1)}&&& \parbox{5cm}{\scalebox{0.5}{\hspace{-0.5cm}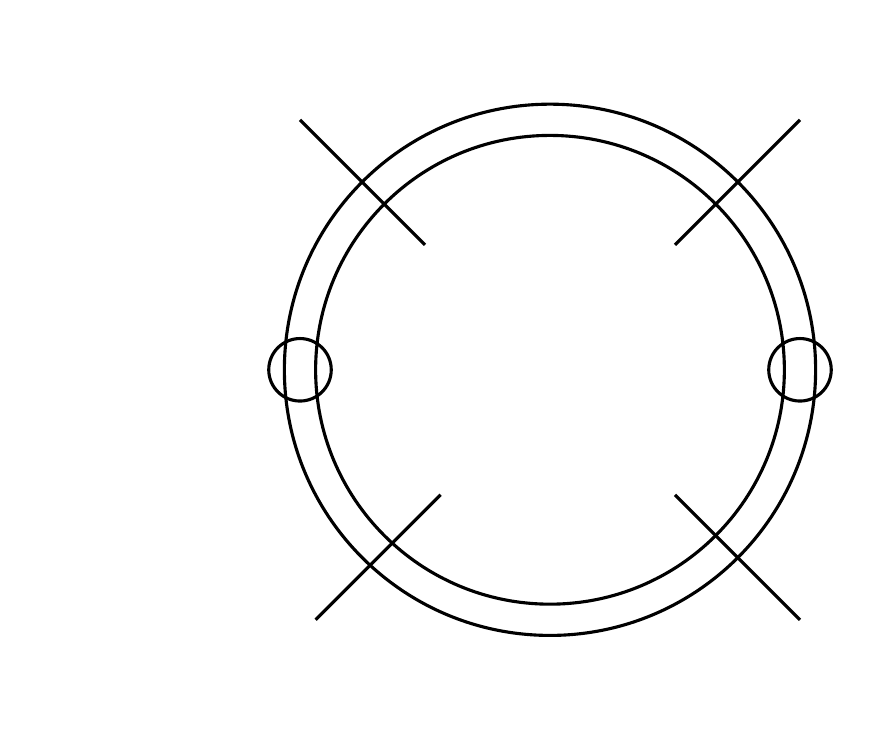}} 
&&  & \parbox{5cm}{\scalebox{0.5}{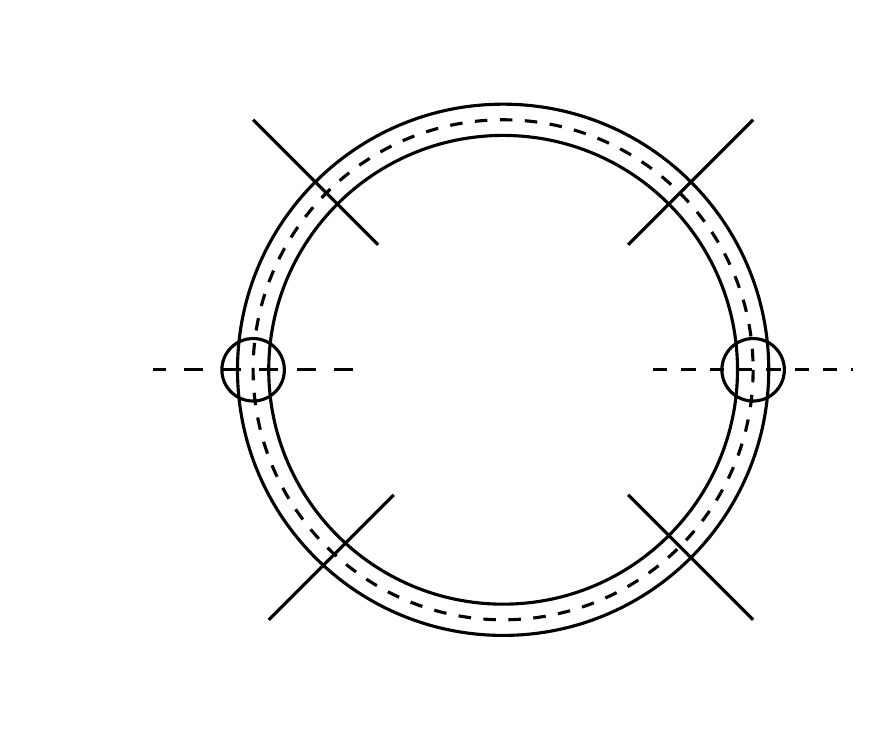}} \notag\end{align}
\caption{The brane configurations of the affine Dynkin quivers. The big circle and dotted circle stand for $N$ D3-branes and one O3-plane, respectively. A segment with numbers $(p,q)$ represents a $(p,q)$ five-brane, and a dotted segment is an O5-plane. The small circles are the orbifold planes, which are always induced in the presence of both O3- and O5-planes.
\label{fig:braneconstruction}
}
\end{figure}

The $A_n^{(1)}$ type $U(N)^{n+1}$ CS theories are the simplest to construct.  
The example $A_3^{(1)}$ is shown in the top left of figure \ref{fig:braneconstruction}.
We place a stack of $N$ D3-branes in the $0123$ directions and periodically identify the 3-direction.  In type IIB string theory there exist so-called $(p,q)$-branes which are 
bound states of $p$ NS5-branes and $q$ D5-branes.  We take $n+1$ $(1,q_a)$-branes and intersect the D3-branes at intervals around the circle in the 3-direction.  The $(1,q_a)$-branes fill the 012 directions.
 To preserve $\calN=3$ SUSY, the 5-branes are tilted at an angle $\theta_a$ in the $45$, $67$, and $89$ planes where  $\theta_a \equiv \arg(1+iq_a)$
\cite{Kitao:1998mf,Bergman:1999na}.    To each D3-brane interval between a $(1,q_a)$- and $(1,q_{a+1})$-branes, we associate a $U(N)$ gauge group.  The CS level is determined by the difference $k_a = q_{a+1} - q_a$.  If required, we can add additional hypermultiplets in the fundamental representations of the $U(N)$ groups by intersecting the D3-branes with D5-branes at angle $\theta = \pi/2$.

To convert the $U(N)$ groups in this $A_n^{(1)}$ quiver into $O/USp$ groups, we add an O3 plane parallel to and coincident with the stack of D3-branes \cite{Feng:2000eq}.  
There are actually four different types of O3 planes which are conventionally denoted $\Opm{3}$ and $\tOpm{3}$.  The D3-branes parallel to $\Om{3}$ and $\tOm{3}$ planes support $O$ type gauge theories, while the D3-branes parallel to $\Op{3}$ and $\tOp{3}$ planes support $USp$ theories.  

The presence of an O3 plane means that type IIB string theory now allows for $\half$ NS5-branes, $\half$ D5-branes, and $\half$ D3-branes provided they intersect the O3 plane.  
Given that the D3-brane charge of an $\Om{3}$ plane is $-1/4$ and the D3-brane charge of an $\tOm{3}$ plane is $+1/4$, one simple way of thinking about an $\tOm{3}$ plane is as a bound state of an $\Om{3}$ plane and a $\half$ D3-brane.  It then makes sense that we associate a $O(2N)$ gauge group with an $\Om{3}$ plane and an $O(2N+1)$ gauge group with an $\tOm{3}$ plane.

Charge conservation arguments imply that the O3 plane changes type when it crosses a $\half$ NS5-brane or $\half$ D5-brane.  
Crossing a $\half$ D5-brane converts an $\Om{3}$ plane to an $\tOm{3}$ plane or an $\Op{3}$ plane to an $\tOp{3}$ plane.  Similarly,
crossing a $\half$ NS5-brane converts an $\Om{3}$ plane to an $\Op{3}$ plane or an $\tOm{3}$ plane to an $\tOp{3}$ plane.  Thus a $\half$ NS5-brane causes the gauge group to change from $O$ to $USp$.  

These rules are summarized in the table (\ref{O3plane}).
We have called the $USp$ group associated with an $\tOp{3}$ plane $USp'(2N)$ to indicate that it will always be neighbored by $O(2N+1)$ groups instead of $O(2N)$ groups.  In other words, the $USp'(2N)$ group has an extra couple of half hypermultiplets compared with the $USp(2N)$ group.

\be
{\renewcommand{\arraystretch}{1.3}
\begin{array}{c|c|c|c|c
}
\text{Type} & \text{D}3 \, \text{charge} & \text{gauge group for }\, N\, \text{D3's} & \frac{1}{2}\,\text{NS5} & \frac{1}{2}\,\text{D}5 
\, \\
\hline
\Om{3} & -\frac{1}{4} & O(2N) & \Op{3} & \tOm{3} 
\\
\tOm{3} & \frac{1}{4} & O(2N+1) & \tOp{3} & \Om{3} 
\\
\Op{3} & \frac{1}{4} & USp(2N) & \Om{3} &  \tOp{3} 
\\
\tOp{3} & \frac{1}{4} & USp'(2N) & \tOm{3} & \Op{3} 
 \\
\end{array}\label{O3plane}
}
\ee

At the level of this brane construction, folding an $A_n^{(1)}$ quiver consisting of $U$ groups to get an $A_n^{(1)}$ quiver consisting of $O/USp$ groups amounts to adding an O3 plane on top of the stack of D3-branes and removing half of the D-branes.
The example $A_3^{(1)}$ is shown in the first row of figure \ref{fig:braneconstruction}.
To remove half the D-branes, some quantities in the unfolded construction must be even.
The total number of NS5-branes $n+1$ (or equivalently the number of gauge groups) must be even so that moving around the circle we can get back to the type of gauge group we started with after having crossed $n+1$ $\half$ NS5-branes.  
Similarly, the total number of D5-branes $\sum_a q_a$ must be even. 
Interestingly, these brane constructions allow the CS levels associated with the individual $O$ and $USp$ gauge groups to be odd and half integral respectively.  The constraint that $\sum_a q_a$ must be even means that there is no overall parity anomaly.

Another type of procedure we can perform on the $A_n^{(1)}$ $U(N)^{n+1}$ quiver is an orientifold or orbifold of the  circle in the 3-direction.  We place orientifold and/or orbifold planes at $x_3 = 0$ and $x_3 = \pi$, assuming the circle has circumference $2 \pi$.  
The orbifold plane $\calI (-1)^{F_L}$ reverses the orientation of the 3-, 5-, 7-, and 9- directions and is thus parallel to an NS5-brane in our construction.  To preserve the same amount of SUSY as the NS5-brane, we also have to include a factor of $(-1)^{F_L}$ acting on the left moving world-sheet fermions.  
There are two types of orientifold planes, $\Op{5}$ and $\Om{5}$, that we will add.\footnote{%
 We ignore the existence of $\tOpm{5}$ planes.
}
To preserve the same amount of SUSY as a D5-brane, they must be placed parallel to the D5-branes in the $012579$ directions, thus reversing the orientation of the $3468$ directions.  The O5$^\pm$ branes have $\pm 1$ unit of D5-brane charge.
The locations of these branes and planes are summarized in table \ref{tb:branes}.

\begin{table}[h]
\centering
\begin{tabular}{c|cccccccccc}
& 0 & 1 & 2 & 3 & 4 & 5 & 6 & 7 & 8 & 9\\
\hline
D3/O3 & $\circ$ & $\circ$ & $\circ$ & $\circ$ & & & & & & \\
D5/O5 & $\circ$ & $\circ$ & $\circ$ &  & & $\circ$ & & $\circ$ & & $\circ$\\
NS5/$\CI (-1)^{F_L}$ & $\circ$ & $\circ$ & $\circ$ &  & $\circ$ & & $\circ$ & & $\circ$ & \\
\end{tabular}
\caption{Orientation of the D-branes, orientifold planes, and orbifold planes.}
\label{tb:branes}
\end{table}

The presence of these orbifold and orientifold planes changes the open string spectrum on the D3-branes and alters the corresponding gauge theory.  Let us start with the orientifold planes.  In analogy to the O3 planes, a stack of $N$ D5-branes coincident with an  $\Om{5}$ plane produces an $O(2N)$ gauge group while the same stack parallel to an $\Op{5}$ brane results in a $USp(2N)$ group.  With respect to these D5-branes, D3-branes look like the addition of flavor fields.  Thus, the gauge theory associated with a stack of $N$ D3-branes is reversed as compared with the D5-branes, $USp(2N)$ for an  $\Om{5}$ plane and $O(2N)$ for an $\Op{5}$ plane \cite{Gimon:1996rq}.  

If we start with an $A_n^{(1)}$ brane construction with a $\mathbb{Z}_2$ reflection symmetry such that each $(1,q_a)$-brane at $x_3 \neq 0$, $\pi$ has a mirror image $(1,-q_a)$-brane at $-x_3$, then 
placing O5 planes on the circle folds an $A_{2n-1}^{(1)}$ type quiver to produce a $C_n^{(1)}$ type quiver where the $O/USp$ gauge groups at the ends depend on the choice of O5 plane.  
The example $C_2^{(1)}$ is shown in the third row of figure \ref{fig:braneconstruction}.
In section \ref{sec:matrixmodels}, 
we discussed the possibility of $O \times O$ and $USp \times USp$ quiver theories for a $C_1^{(1)}$ Dynkin diagram.  
We note in passing that we can realize these quivers by placing a single $(1,q)$ brane between either
two $\Op{5}$ or two $\Om{5}$ planes respectively.  
Similar types of orientifold constructions are discussed in detail in \cite{Uranga:1998uj,Hanany:1997gh} for 3+1 and 5+1 dimensional cousins of these quiver theories.\footnote{%
 Note that if a 
 1/2 NS5-brane is coincident with the orientifold plane, then the folding prescription is modified.  
 The $A_n$ quiver is folded on an edge rather than a node.  
 The gauge group at the end of the quiver will be $U$ type rather than $O/USp$, and there will 
 be a  matter field in a two index representation of the $U$ group at the end of the quiver, antisymmetric 
 for $\Om{5}$ and symmetric for $\Op{5}$ \cite{Hanany:1997gh}.
 \label{matterfootnote}
}

Adding an orbifold plane produces a forked end to the quiver.  
In the presence of the orbifold plane $\calI(-1)^{F_L}$, there are two types of D3-brane which we could call D3$^\pm$ \cite{Kapustin:1998fa, Sen:1998ii}.  If we start with an $A_n^{(1)}$ $U(2N)^{n+1}$ quiver with equal types of D3$^+$ and D3$^-$ branes, after adding an orbifold plane, the $U(2N)$ group from the D3-branes next to the orbifold plane will break to $U(N) \times U(N)$.  
If we include two orbifold planes on the circle, we fold an $A_{2n-5}^{(1)}$ type quiver down to a $D_n^{(1)}$ quiver. 
 We can also contemplate more elaborate constructions where we include an orbifold plane at one end of the stack of D3-branes and an orientifold at the other, thus folding an $A_{2n-3}^{(1)}$ quiver down to an 
 $A_{2n-1}^{(2)}$ quiver.  The examples $A_5^{(2)}$ and $D_4^{(1)}$ are shown in the second and fourth rows of figure \ref{fig:braneconstruction}.
 
We also have to worry about the CS levels.  The CS levels of the $U$ type groups in the middle of the quiver are defined exactly as they were before, as the difference $q_{a+1} - q_a$ in the 5-brane charges.  
An $O/USp$ group or a $U(N)^2$ factor at the end of the quiver descends from a $U$ group sandwiched between a $(1,q)$ and a $(1,-q)$ brane to which we associate a CS level $2q$.  Reassuringly, given 
the unfolding relations (\ref{vONunfold}), (\ref{vUSpNunfold}) and (\ref{vDnunfold}), $2q$ is even.

Finally, one can envision combining these three elements: orbifold planes, O5 planes and O3 planes. 
It turns out that the product of the space-time and world-sheet
reflection symmetries of two of these objects is the reflection symmetry generated by the third.  Thus, if two of these objects are present, the third one is present as well.  
Consulting ref.\ \cite{Hanany:1997gh}, and starting with a stack of $2N$ D3-branes, 
we find that intersecting an $\Opm{5}$ plane with an 
$\Opm{3}$ plane will produce a $U(N)$ group living on the D3-branes between the O5 plane and the nearest $(1,q)$-brane.
In contrast, intersecting an $\Om{5}$ plane with an 
$\Op{3}$ plane will yield a $USp(N_1) \times USp(N_2)$ group where $N_1+N_2=2N$ and $N_1$ and $N_2$ must be even.
Switching the charges on the orientifold planes, we can replace $USp(N_1) \times USp(N_2)$ with
$O(N_1) \times O(N_2)$.
In sum, by suitably adjusting the charges on the orientifold planes, we can construct the $B_n^{(1)}$ quivers,
the $O/USp$ version of the $D_n^{(1)}$ quivers, and 
 the $D_{n}^{(2)}$ quivers.  (In these constructions, we still have the restriction that the CS levels of the gauge groups at a forked end of the quiver be equal.)  Examples of such constructions are shown in the
 second column of figure \ref{fig:braneconstruction}.

In section \ref{ss:unfold}, we discussed some extra factors of $\sinh [2 \pi \lambda_l]$ that occur in unfolding the matrix model that contribute (at large $N$) in the same way as fundamental fields would. 
We now explain how these factor are connected with the brane constructions.
Since the O5 planes carry D5-brane charge, the relation between the $A_{2n-1}^{(1)}$ quiver and the $C_n^{(1)}$ quiver involves some extra flavor fields.  An $\Op{5}$ plane lifts to a theory with effectively one additional D5-brane or fundamental hypermultiplet field.  An $\Om{5}$ plane lifts to a theory with one anti-D5-brane which one could think of as reducing by one the total number of fundamental hypermultiplets.
Reassuringly, this counting is consistent with the extra factors of $\sinh [2 \pi \lambda_l]$ in  (\ref{vONunfold}) and (\ref{vUSpNunfold}).
More complicated examples with orbifold planes work similarly.

%%%%%%%%%%%%%%%%%%%%%%%%%%%%%%%%%%%%%
\section{Examples}
\label{sec:examples}
%%%%%%%%%%%%%%%%%%%%%%%%%%%%%%%%%%%%%

%%%%%%%%%%%%%%%%%%%%%%%%%%%%%%%%%%%%%
\subsection{The $O(2N)_{2k}\times USp(2N)_{-k}$ theories}
%%%%%%%%%%%%%%%%%%%%%%%%%%%%%%%%%%%%%
As a particular example of the brane constructions of section \ref{sec:branes}, we can realize
an $O(2N)_{2k}\times USp(2N)_{-k}$ theory \cite{Hosomichi:2008jb, Aharony:2008gk}
as an orientifold projection of the
$U(2N)_{2k} \times U(2N)_{-2k}$ ABJM model \cite{Aharony:2008ug}.
The $U(N)_{k} \times U(N)_{-k}$ ABJM model is an $\CN = 6$ supersymmetric 
Chern-Simons-matter theory realized by $N$ D3-branes wrapped on a circle
with an intersecting NS5-brane and a $(1,k)$ 5-brane in type IIB string theory. 
After a T-duality, this brane construction lifts to $N$ coincident M2-branes in M-theory sitting at the 
origin of the orbifold space $\mathbb{C}^4/\BZ_k$.
There are four bifundamental (chiral) fields $(A_1, A_2, B_1^\ast, B_2^\ast)$ in the theory, and
the supersymmetry enhances to $\CN =8$ when $k=1,2$.

As discussed in section \ref{sec:branes}, the orientifolded  $O(2N)_{2k} \times USp(2N)_{-k}$ theory
can be realized
by adding an O3 plane to the type IIB brane construction.
The orientifolded theory has $\CN=5$
SUSY \cite{Hosomichi:2008jb,Aharony:2008gk}.
In terms of the field theory, the orientifold projection acts on the four bifundamental fields of the 
ABJM model $(A_1,A_2,B_1^*,B_2^*)$ with $U(2N)\times U(2N)$ gauge group as
\begin{align}\label{orientifoldproj}
	A_1 = B_1^T J \ , \qquad A_2 = B_2^T J \ ,
\end{align}
where $J$ is the invariant anti-symmetric matrix of the $USp(2N)$ group acting 
from right on the anti-fundamental indices of $U(2N)$.
The resulting model has two bifundamental half hypermultiplets $(A_1,A_2)$
with $O(2N)\times USp(2N)$ gauge symmetry.

The leading order effect of the O3 plane on the moduli space of the ABJM theory 
is to quotient by a $\mathbb{Z}_2$ group.
The moduli space of a single M2-brane in 
the orientifolded theory is $\BC^4/\hat {\bf D}_k$ where $\hat {\bf D}_k$ is the
binary dihedral group with $4k$ elements.
The simplest case is $\hat {\bf D}_1 = \BZ_4$, which implies the equivalence 
between the $O(2N)_{\pm 2}\times USp(2N)_{\mp 1}$ CS-matter theory and
the ABJM model with $U(N)_{4} \times U(N)_{-4}$ gauge groups \cite{Aharony:2008gk}.
As discussed in the introduction, in general the M2-brane 
moduli space of these ${\mathcal N}=3$ matter CS theories is
a four complex dimensional hyperk\"ahler cone whose level surface is a seven real dimensional tri-Sasaki Einstein space $Y$.  Through the relation between the free energy and $\Vol(Y)$ (\ref{GRfree}), we can
check this expectation that adding an orientifold plane quotients the moduli space and halves the volume of $Y$.

The matrix model of the $O(2N)_{2k}\times USp(2N)_{- k}$ theory
consists of the vector multiplets for $O(2N)$ and $USp(2N)$ gauge groups 
and two bifundamental fields between them:
\begin{align}\label{ABJ}
	Z = \frac{1}{(N!)^2} \int &\left( \prod_{i,j=1}^N d \lambda_i d \mu_j \right)
e^{2\pi i k \sum_{i} (\lambda_i^2 - \mu_i^2)} \,
\frac{ \left(\prod_{i<j} 4 \sinh[ \pi (\lambda_i - \lambda_j)] \sinh[\pi(\mu_i - \mu_j)] \right)^2}
{\left(\prod_{i,j} 2 \cosh[ \pi (\lambda_i - \mu_j)] \right)^2 } \nonumber \\
&
\times
 \frac{ \left(\prod_{i < j} 4 \sinh[ \pi (\lambda_i + \lambda_j)] \sinh[\pi(\mu_i + \mu_j)] \right)^2}
{\left(\prod_{i,j} 2 \cosh[ \pi (\lambda_i + \mu_j)] \right)^2 }
 \left( \prod_i 2 \sinh [2 \pi \lambda_i] \right)^{2} \ .
\end{align}
Now we demonstrate how the unfolding trick works.
Let $\ell_I$ and $m_J$ be
$\ell_i = \lambda_i$, $m_i = \mu_i$ and $\ell_{i+N} = - \lambda_i$ and $m_{i+N} = - \mu_i$, respectively. 
Then, the matrix model \eqref{ABJ} becomes 
\begin{align}
	Z = \frac{1}{(N!)^2} \int &\left( \prod_{I=1}^{2N} d \ell_I d m_I \right)
e^{2\pi i k \sum_I (  \ell_I^2 -m_I^2) / 2} 
 \frac{ \left( \prod_{I<J} 4 \sinh[ \pi (\ell_I - \ell_J)] \sinh[\pi (m_I - m_J)] \right)}
{\left( \prod_{I,J} 2 \cosh[\pi (\ell_I - m_J) ] \right) } \nonumber \\
&\times \left( \prod_J \frac{  \sinh [2 \pi \ell_J] }{\sinh [2 \pi m_J] } \right) \ .
\end{align}
The integrand in the first line is the square root of the integrand of 
the matrix model of the $U(2N)_{2k}\times U(2N)_{-2k}$ ABJM theory.
The integrand in the second line will not contribute in the 
large $N$ limit.
The remaining expression 
has a saddle point very similar to that of the partition function of the ABJM theory.  If 
we let $F_{ABJM}(N,k)$ be the free energy of the ABJM $U(N)_{k}\times U(N)_{-k}$ theory, we find that our 
free energy must be $F = F_{ABJM}(2N,2k) / 2$.  As the free energy scales as $k^{1/2}N^{3/2}$, it follows that in the large $N$ limit $F = \sqrt{2} F_{ABJM}(N,2k)$.  The volume of the tri-Sasaki Einstein space is proportional to $1/F^2$.  Thus our volume must be
\begin{align}\label{vol}
	\text{Vol}(Y) = \text{Vol}(Y_{ABJM,\, (N,2k)})/2 \ ,
\end{align}
with the expected factor of two.
Similarly, one can check that the free energy of the $O(2N)_{\pm 2}\times USp(2N)_{\mp 1}$ theory is the same as that of the $U(N)_{4} \times U(N)_{-4}$ 
ABJM theory in the large $N$ limit.

%%%%%%%%%%%%%%%%%%%%%%%%%%%%%%%%%%%%%
\subsection{Unfolding $D_4^{(1)}$ and folding $A_3^{(1)}$}
%%%%%%%%%%%%%%%%%%%%%%%%%%%%%%%%%%%%%
In the previous section, the unfolding trick 
was used to change the gauge groups from 
orthosymplectic to unitary without changing the quiver.
Here we present a different illustration of unfolding where the type of quiver is changed: 
We unfold the $D_4^{(1)}$ quiver to get an $A_3^{(1)}$ quiver and then fold $A_3^{(1)}$ to get $C_2^{(1)}$.

\begin{figure}[h]
\begin{align*}
D_4^{(1)}: && \parbox{5cm}{\scalebox{0.45}{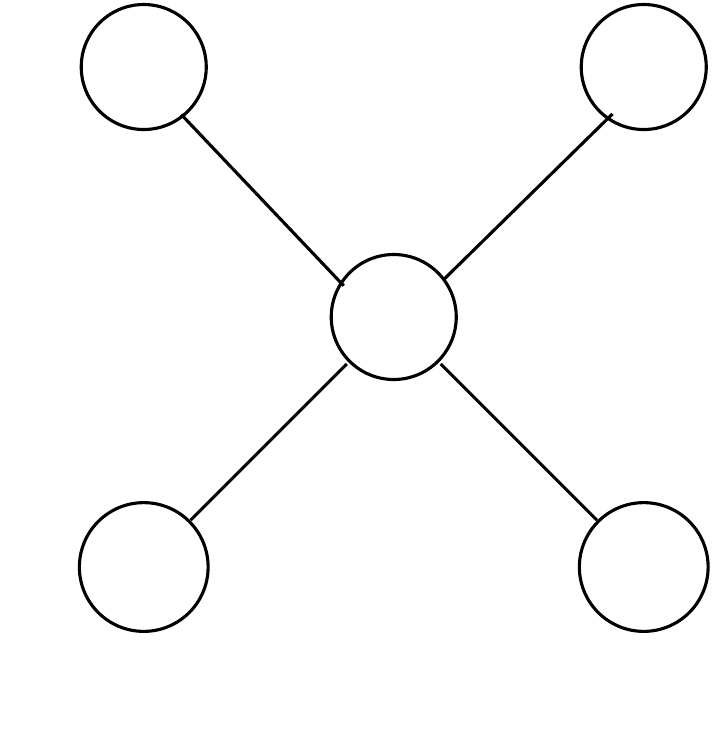}}  
A_3^{(1)}: && \parbox{5cm}{\scalebox{0.45}{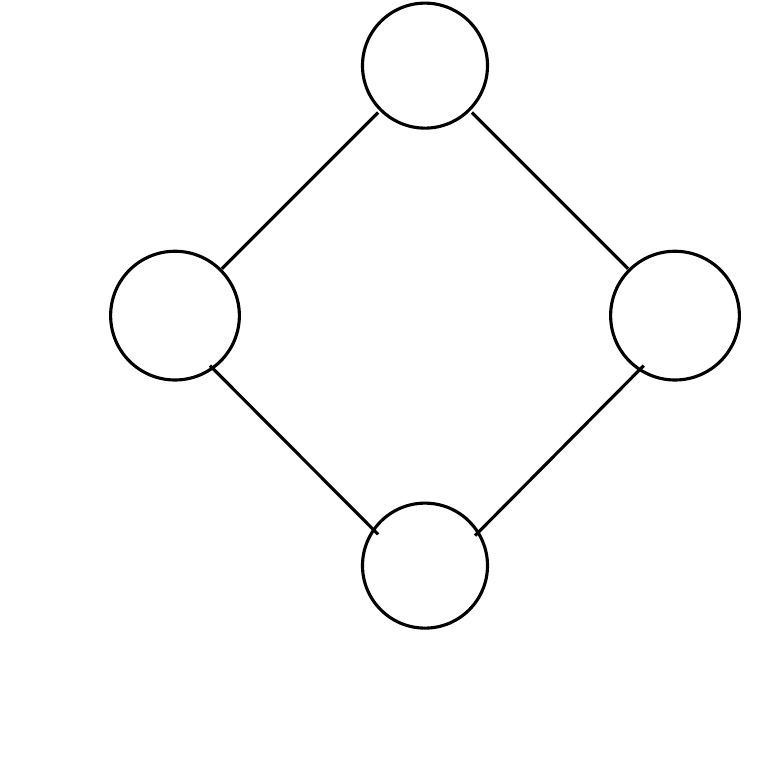}} 
C_2^{(1)}: & \quad \parbox{5cm}{\scalebox{0.45}{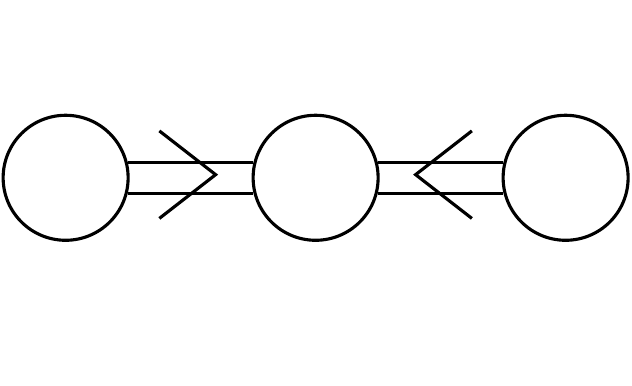}}
\end{align*}
\caption{
\label{fig:D4A3}
The  $D_4^{(1)}$, $A_3^{(1)}$ and $C_2^{(1)}$ quivers are related by the unfolding trick. The numbers in and next to the circles denote the rank of the gauge groups and the Chern-Simons levels, respectively.}
\end{figure}

The $D_4^{(1)}$ quiver is shown in figure \ref{fig:D4A3}, where the ranks (divided by $N$) of the unitary groups, the CS levels and the eigenvalues are given in, next to and below the nodes, respectively.
The partition function of the $D_4^{(1)}$ quiver takes the following form:
\begin{align}\label{D4PF}
	Z_{D_4^{(1)}} = \int  \, &L_{V} (U(N),k,\mu) L_{V} (U(N),k,\mu') \cdot L_{V} (U(N),k',\tilde\mu)L_{V} (U(N),k',\tilde\mu') \notag \\
	& \cdot L_{V} (U(2N),k'',\nu) \notag \\
	&\cdot L_M(U(N)\times U(2N), \mu, \nu) L_M(U(N)\times U(2N), \mu', \nu) \notag \\
	& \cdot L_M(U(N)\times U(2N), \tilde\mu, \nu) L_M(U(N)\times U(2N), \tilde\mu', \nu) \ ,
\end{align}
where we have omitted the measure of the eigenvalues for simplicity.

By using the formula \eqref{vDnunfold}, the two $U(N)$ groups of the $D_4^{(1)}$ quiver 
with the eigenvalues and the CS levels $(k,\mu)$ and $(k,\mu')$ are mapped to (the square root of) the $U(2N)$ group with $(2k,\lambda)$. The other two $U(N)$'s with $(k',\tilde\mu)$ and $(k',\tilde\mu')$ are also combined into another $U(2N)$ with $(2k',\tilde\lambda)$.
For the $U(2N)$ of $D_4^{(1)}$ with $(k'',\nu)$, we introduce two eigenvalue sets $\kappa=\nu$ and $\tilde\kappa=\nu$ and rewrite the partition function of the vector multiplet
\begin{align}
	L_V(U(2N),k'',\nu) = L_V(U(2N),k'',\kappa)^{1/2} L_V(U(2N),k'',\tilde\kappa)^{1/2}  \ ,
\end{align}
where the eigenvalue distributions for $\kappa$ and $\tilde\kappa$ are supposed to be equal by definition.
The formula \eqref{vDnunfold} lets us write the four bifundamentals in $U(N)\times U(2N)$ of $D_4^{(1)}$ as (the square root of) four bifundamentals in $U(2N)\times U(2N)$.
All of these operations lead to the following expression of the $D_4^{(1)}$ partition function \eqref{D4PF}
\begin{align}\label{A3PF}
	Z_{D_4^{(1)}} = \int & L_V(U(2N),2k,\lambda)^{1/2} L_V(U(2N),2k',\tilde\lambda)^{1/2} \notag \\
	& \cdot L_V(U(2N),k'',\kappa)^{1/2} L_V(U(2N),k'',\tilde\kappa)^{1/2} \notag \\
	& \cdot L_M(U(2N)\times U(2N), \lambda, \kappa)^{1/2} L_M(U(2N)\times U(2N), \lambda, \tilde\kappa)^{1/2} \notag \\
	& \cdot L_M(U(2N)\times U(2N), \tilde\lambda, \kappa)^{1/2} L_M(U(2N)\times U(2N), \tilde \lambda, \tilde \kappa)^{1/2} \ .
\end{align}
Here the eigenvalue distributions of $\kappa$ and $\tilde\kappa$ are assumed to be equal. The integrand of \eqref{A3PF} is the square root of that of the $A_3^{(1)}$ quiver shown in figure \ref{fig:D4A3}. 
The saddle point equations of (\ref{A3PF}) and $A_3^{(1)}$ 
give the same eigenvalue distribution.

We can now fold this $A_3^{(1)}$ quiver to obtain a $C_2^{(1)}$ quiver of the form
$O(2N)_{2k} \times U(2N)_{k''} \times USp(2N)_{k'}$.  Let us label the eigenvalues of the $O$, $U$ and $USp$ groups $\alpha$, $\beta$, and $\gamma$ respectively.  
Making the appropriate identifications between the $A_3^{(1)}$ eigenvalues $(\lambda, \lambda', \kappa, \kappa')$ and the $C_2^{(1)}$ eigenvalues $(\alpha, \beta, \gamma)$, 
we find that
\begin{align}
Z_{C_2^{(1)}} = Z_{D_4^{(1)}} =& \int L_V(O(2N), 2k, \alpha) L_V(U(2N), k'', \beta) L_V(USp(2N), k', \gamma)
\nonumber
\\
&
\hspace{5mm}
\cdot L_M(O(2N)\times U(2N), \alpha, \beta) L_M(U(2N) \times USp(2N), \beta, \gamma) \ .
\end{align}
Here we used the formulae \eqref{vONunfold} and \eqref{vUSpNunfold} for the nodes
at both ends and the formula \eqref{vectorunfold} for the middle node of $C_2^{(1)}$.

%%%%%%%%%%%%%%%%%%%%%%%%%%%%%%%%%%%%%
\section{The $\widehat{\text{ADE}}$ quiver matrix models and generalized Seiberg duality}
\label{sec:ADEmatrix}
%%%%%%%%%%%%%%%%%%%%%%%%%%%%%%%%%%%%%

In this section, we relate our affine ADE matrix models for $\calN=3$ CS matter theories with unitary groups
to a class of conformal multi-matrix models first introduced by refs.\ \cite{Kharchev:1992iv,Kostov:1992ie}.  We call these multi-matrix models $\widehat{\text{ADE}}$ quiver matrix models.
It turns out that these same $\widehat{\text{ADE}}$ quiver matrix models
were used to study $\calN=1$ ADE quiver theories in 3+1 dimensions
\cite{Cachazo:2001gh,Cachazo:2001sg}.  One interesting consequence of this relationship is that at the level of the matrix model
Seiberg duality acting on the 3+1 dimensional gauge theory is the same as a 2+1 dimensional analog of Seiberg duality \cite{Aharony:1997gp,Giveon:2008zn} acting on the CS theory.
In both cases, the duality is equivalent to the action of the Weyl group associated with the ADE Dynkin diagram.

The partition function for the $\widehat{\text{ADE}}$ quiver matrix models is given by
\begin{align}\label{ADEMM}
	Z = \int \left( \prod_{a=1}^{r} \prod_{i=1}^{N_a} d\lambda_{a,i}\right) \, e^{-\sum_{a=1}^r
	 \sum_{i=1}^{N_a} W_a(\lambda_{a,i})} 	\prod_{a}\prod_{i<j} (\lambda_{a,i} - \lambda_{a,j})^2
	\prod_{a<b}\prod_{i, j} (\lambda_{a,i} + \lambda_{b,j})^{(\alpha_a,\alpha_b)} \ ,
\end{align}
where $r$ is the rank of the Dynkin diagram and $\alpha_a$ 
are the simple roots.
The roots are normalized such that $(\alpha_a,\alpha_a) = 2$ and $(\alpha_a,\alpha_b)=-1$
for the adjacent nodes. Note that the eigenvalues $\lambda_{a,i}$ must be 
positive \cite{Kostov:1992ie,Kostov:1999xi}.

To relate this model to ours, we redefine the eigenvalues as follows:
\begin{align}
	\lambda_{a,i} \to e^{2\pi \lambda_{a,i}} \ ,
\end{align}
where the new eigenvalues $\lambda_{a,i}$ run from $-\infty$ to $\infty$.
After this redefinition, the matrix model \eqref{ADEMM} becomes 
\begin{align}
\label{Ztransform}
	Z = \int &\left(\prod_{a=1}^{r} \prod_{i=1}^{N_a} d\lambda_{a,i} \right)\,  
	e^{\pi \sum_{a} (\sum_b (\alpha_a, \alpha_b) N_b) \sum_i \lambda_{a,i}}\notag\\
	&\cdot e^{-\sum_{a=1}^r
	 \sum_{i=1}^{N_a} W_a(e^{2\pi\lambda_{a,i}})} \prod_{a}\prod_{i<j} (2\sinh [\pi (\lambda_{a,i} - \lambda_{a,j})])^2
	\prod_{a<b}\prod_{i, j} (2\cosh [\pi (\lambda_{a,i} - \lambda_{b,j})])^{(\alpha_a,\alpha_b)} \ ,
\end{align}
and it looks like the $\CN\ge 3$ matrix models on $S^3$ with unitary gauge groups.
The condition \eqref{nf2nc} is translated to
\begin{align}
	\sum_b (\alpha_a, \alpha_b) N_b = 0 \ ,
\end{align}
which cancels the exponential term in the first line.
If we take the potential terms as
\begin{align}\label{CSpot}
	W_a(x) = - \frac{i k_a}{4\pi} (\log x)^2 \ ,
\end{align}
they reproduce the Chern-Simons terms of the $\CN \ge 3$ matrix models.\footnote{%
 Interestingly, the $\hat A$-type (or necklace quiver) theories 
can be reformulated in terms of an ideal Fermi gas
\cite{Kostov:1995xw}.
}

Modulo some subtle convergence issues, refs.\ \cite{Willett:2011gp,Benini:2011mf} demonstrated that the absolute value 
$|Z_{S^3}|$ of the CS matter theory partition function is invariant under a generalized Seiberg duality
\cite{Aharony:1997gp,Giveon:2008zn}.  
Ref.\  \cite{Gulotta:2011vp} studied this invariance in the large $N$ limit 
for our CS $\calN = 3$ $\otimes_a U(N_a)$ 
quiver theories satisfying $\sum_a n_a k_a = 0$ and (\ref{nf2nc}).\footnote{%
 See also \cite{Amariti:2011uw}.
}
Under a duality
at node $a$, the ranks of the gauge groups do not change but the CS levels and averaged saddle point eigenvalue distributions shift according to the rules
\begin{align}
	k_b &\to k_b - (\alpha_a,\alpha_b)k_a \ , \notag \\
	\sum_{I=1}^{n_a} y_{a,I} &\to \sum_{I=1}^{n_a} y_{a,I} -\sum_{c}\sum_{J=1}^{n_c} (\alpha_a,\alpha_c) y_{c,J} \ .
\end{align}

Given the relation between these $\otimes_a U(N_a)$ 
quiver theories and the $\widehat{\text{ADE}}$ quiver matrix models, we note that this type of Seiberg duality
is a special case of a more general Seiberg duality that acts on a matrix model with 
arbitrary potential terms $W_a(x)$ \cite{Cachazo:2001sg,Dijkgraaf:2003xk}.  
Under this more general Seiberg duality, again at node $a$,
\begin{align}
N_a &\to N_a - \sum_{b} (\alpha_a,\alpha_b)N_b \ , \\
W_b (\lambda_{b}) &\to W_b (\lambda_{b}) - (\alpha_a, \alpha_b) W_a (\lambda_{b}) \ . \notag 
\end{align}
In our case, $N_a$ does not change under Seiberg duality due to the 
condition \eqref{nf2nc}.  
For the affine ADE quiver gauge theories in 
four dimensions, there is the constraint on the potentials $\sum_a N_a W_a = 0$ \cite{Cachazo:2001gh,Cachazo:2001sg}.
Although it was originally derived from geometric construction of the super potentials in four dimensions, we may well impose it for our case.
Combined with \eqref{CSpot}, it leads to the constraint on the CS levels $\sum_a n_a k_a=0$ necessary to obtain an M-theory dual.
As pointed out in \cite{Cachazo:2001sg,Dijkgraaf:2003xk}, the invariance under Seiberg duality
is a result of the Weyl reflection symmetry of the Dynkin diagrams 
which acts on
the roots as $\alpha_b  \to \alpha_b - (\alpha_b,\alpha_a) \alpha_a$.
Seiberg duality of the orthosymplectic case in a large $N$ limit can be obtained by using the unfolding trick
\eqref{vONunfold} and \eqref{vUSpNunfold} which relates the matrix models with orthosymplectic groups to those of unitary groups.

%%%%%%%%%%%%%%%%%%%%%%%%%%%%%%
\section{Discussion}
%%%%%%%%%%%%%%%%%%%%%%%%%%%%%%
The principal result of this paper is a classification of a certain type of $\calN=3$ CS matter theory.  Consider theories that consist of a product of the classical groups 
\[
[\otimes_a U(N_a)] \otimes [\otimes_b O(N_b)] \otimes  [\otimes_c  USp(2N_c)] \ ,
\]
with matter fields in bifundamental and fundamental representations.  The form of the superpotential and K\"ahler potential are fixed by the $\calN=3$ SUSY.  Having fixed the gauge groups and the matter content, the only degrees of freedom left are the CS levels $k_a$, which we assume to be nonzero.  
We then compute the partition function of these matrix models in a large $N$ limit
on an $S^3$ using the matrix model of 
ref.\ \cite{Kapustin:2009kz}.
The classification result is that such theories for which the long range forces between the eigenvalues vanish in a saddle point approximation of the matrix model (given the ansatz (\ref{eigenvalueansatz})) have gauge group and bifundamental field content  specified by the Dynkin diagrams in figure \ref{fig:ADEquivers} and the left hand column of figure \ref{fig:twistedADEquivers}.

We offer two speculations about the importance of this result.  The first speculation concerns the AdS/CFT correspondence.  We wonder if there exists some sense in which this
 set of CS matter theories is a
complete classification of conformal 
$\calN=3$ CS matter theories with eleven dimensional 
(and massive type IIA) supergravity duals.  
At least this class of CS theories is one for which we can check that the $S^3$ matrix model free energy agrees with the corresponding gravity calculation.  
Also, the examples in this paper give us a better understanding of what types of CS theories appear in AdS/CFT correspondences, but ideally we would like to say something stronger.

We mention three obstacles in the way of making a stronger statement.  First, the eigenvalue cancellation appears to depend in a nontrivial way on the choice of ansatz (\ref{eigenvalueansatz}).  Perhaps a smarter or more general ansatz would allow for a more general class of theories.  Next, we should clearly allow for more general types of matter fields.  We mentioned in passing in footnote \ref{matterfootnote}  that folding a quiver along an edge instead of a node gives rise to matter fields in symmetric and antisymmetric two index representations.  However, there is nothing to stop us from including matter fields in more exotic representations of the classical groups.  
Finally, there may be gauge theories for which we do not even know how to write the Lagrangian.  For example, taking the S-dual of the type IIB brane configurations described in section \ref{sec:branes} produces $(p,q)$ five-branes where $p>1$.  By S-duality, we know such configurations have a gauge theory description, but we do not know how to write the Lagrangian down directly.  Optimistically, it might still be true that given an arbitrary conformal $\calN=3$ CS matter theory with an 11 dimensional supergravity dual, such a theory will be related by duality to one of the theories in our list.

Related to these last two obstacles is the observation that there are other affine Dynkin diagrams
shown in the left of figure \ref{fig:twistedDynkin} whose interpretation in terms of the gauge theory
is less clear.  Here are three questions we would like to be able to address eventually.  Should these Dynkin diagrams have an interpretation in terms of $\calN=3$ CS matter theories?  If they do have such an interpretation, can their Lagrangians be constructed using matter fields in more exotic representations?  If there is no Lagrangian description, can we relate them to more familiar theories using duality?

The second speculation is mathematical in nature.
Given the constraint on the CS levels (which for the unfolded quivers is that $\sum_a k_a N_a=0$), 
a branch of the moduli space of these
$\calN=3$ theories is an $N$ fold symmetric product of an eight real dimensional hyperk\"ahler cone 
\cite{Jafferis:2008qz}.  The base of such a cone is the tri-Sasaki space $Y$ that appears in
eleven dimensional supergravity.  An open problem is the classification of such seven dimensional tri-Sasaki spaces \cite{Boyer:1998sf}.  We wonder if the list of $\calN=3$ CS matter theories in this paper might provide (or help provide) such a classification.  
An immediate objection is that a cone realized as a hyperk\"ahler quotient 
always has a base which is tri-Sasaki which in turn could be used in an eleven dimensional supergravity construction.  Our list of Dynkin diagrams gives only a small subset of such quotient constructions.  Additionally, it seems likely that there exist hyperk\"ahler cones which are not 
hyperk\"ahler quotients.  
Optimistically, if our list is in some sense complete for $\calN=3$ CS matter theories and if the AdS/CFT correspondence is correct, then our list must also be complete for tri-Sasaki spaces.
While this mathematical speculation may be too ambitious, we would still like to understand from a purely geometric point of view what is special about our set of tri-Sasaki spaces.

In conclusion, we would like to mention some aspects of this paper that should be developed further.
First, it would be nice to understand how the Lie group associated with the Dynkin diagram appears in the physics of our $\calN=3$ CS matter theories.  Clearly the Weyl group is important in understanding Seiberg duality.  Also, we know from ref.\ \cite{Gaiotto:2008ak} that if we set the CS levels to zero which will enhance the SUSY to $\calN=4$, the Coulomb branch of the moduli space has an enhanced global symmetry given by the corresponding simple Lie group.  
In more detail  there is a topological $U(1)$ global symmetry for 
each node whose current is given by $J_a = \frac{1}{4\pi} \ast F_a$, where $F_a$ is
the field strength of the gauge theory of the node labeled by $a$.
These $U(1)$ symmetries give rise to the Cartan subgroups, and enhance to
the global symmetry of the quiver diagram when 
combined with the monopole operators.
However, it is not clear to us how nonzero CS levels affect this global symmetry enhancement.

Second, we have been brief in our treatment of the parity of the gauge group ranks and the CS levels.  The D-brane construction of section \ref{sec:branes} suggests $USp$ and $O$ groups can have half integral and odd CS levels respectively in these quiver constructions, 
despite the standard argument to the contrary 
reviewed in appendix \ref{app:CSnorm}.  We only briefly discussed
how $\tOpm{3}$ planes allow for $O(2N+1)$ and $USp'(2N)$ groups, and we neglected 
$\tOpm{5}$ planes which presumably add extra fundamental hypermultiplets to the quiver theory.

Third, we expect
that in a manner similar to the Alday-Gaiotto-Tachikawa conjecture \cite{Alday:2009aq}, the quiver gauge theories we considered will have a connection to Toda field theory.
The $\widehat{\text{ADE}}$ quiver matrix models we introduced in section \ref{sec:ADEmatrix} make
clear a relation between generalized Seiberg duality in our three dimensional CS matter theories and 
Seiberg duality of the $\widehat{\text{ADE}}$ 
quiver gauge theories in four dimensions.
Without the CS levels, the partition function of the $\hat A_n$ quiver matrix models can be written
as a correlation function of vertex operators of the $c=n$ Toda field theory \cite{Kostov:1995xw,Dijkgraaf:2009pc}. Although the Stieltjes-Wigert type potential is not straightforward to include, we expect
that the quiver gauge theories we considered have some connection to Toda field theory.

%%%%%%%%%%%%%%%%%%%%%%%%%%%%%%%%%%%%
\section*{Acknowledgments} We would like to thank A.~Hanany, P.~Putrov, M.~Ro\v{c}ek, S.~Sugimoto, and Y.~Tachikawa for discussion. This work was supported in part by the US NSF under Grants No.\,PHY-0844827 and PHY-0756966.  CH thanks the Sloan Foundation for partial support. TN thanks IPMU for hospitality during the completion of this paper.

%%%%%%%%%%%%%%%%%%%%%%%%%%%%%%%%%%%%
\appendix
%%%%%%%%%%%%%%%%%%%%%%%%%%%%%%%%%%%%
\section{Adjoint and fundamental representations of simply laced Lie groups}
\label{app:reps}
%%%%%%%%%%%%%%%%%%%%%%%%%%%%%%%%%%%%

\subsubsection*{SU($N$)}

Let ${\bf e}_i$ be unit vectors in ${\mathbb R}^N$ with the inner product ${\bf e}_i \cdot {\bf e}_j = \delta_{ij} - 1/N$.  
The fundamental representation has the $N$ weights ${\bf e}_i$, $i=1, \ldots, N$. 
The adjoint representation has the $N(N-1)$ weights (or roots) ${\bf e}_i - {\bf e}_j$, $i \neq j$, along with $N-1$ elements that generate the Cartan sub-algebra.

\subsubsection*{SO($2N$)}

Let ${\bf e}_i$ be unit vectors in ${\mathbb R}^N$ with the standard inner product.  
The fundamental representation has the $2N$ weights $\pm {\bf e}_i$, $i=1, \ldots, N$. 
The adjoint representation has the $2N(N-1)$ roots ${\bf e}_i \pm {\bf e}_j$, $i \neq j$ along with $N$ elements that generate the Cartan sub-algebra.  

\subsubsection*{SO($2N+1$)}

Let ${\bf e}_i$ be unit vectors in ${\mathbb R}^N$ with the standard inner product.  
The fundamental representation has the $2N$ weights $\pm {\bf e}_i$, $i=1, \ldots, N$ along with $0$. 
The adjoint representation has the $2N(N-1)$ roots ${\bf e}_i \pm {\bf e}_j$, $i \neq j$ along with 
$2N$ roots $\pm {\bf e}_i$ and $N$ elements that generate the Cartan sub-algebra.  

\subsubsection*{USp($2N$)}

Let ${\bf e}_i$ be unit vectors in ${\mathbb R}^N$ with the standard inner product.  
The fundamental representation has the $2N$ weights $\pm {\bf e}_i / \sqrt{2}$, $i=1, \ldots, N$. 
The adjoint representation has the $2N(N-1)$ roots $({\bf e}_i \pm {\bf e}_j)/\sqrt{2}$, 
$i \neq j$ along with 
$2N$ roots $\pm \sqrt{2} {\bf e}_i$ and $N$ elements that generate the Cartan sub-algebra.  Note we have included factors of $\sqrt{2}$ so that the longest weights will have their conventional length squared equal to two.

\vskip 0.2in

\noindent
The ranks of the Weyl groups are
\be
\begin{array}{c|c}
G_a & \operatorname{rank}(\calW_a) \\
\hline
SU(N) & N! \\
O(2N+1) & 2^N N! \\
USp(2N) & 2^N N! \\
O(2N) & 2^{N-1} N! 
\end{array}
\ee

%%%%%%%%%%%%%%%%%%%%%%%%%%%%%%%%%%%%
\section{Normalization of the Chern-Simons action}
\label{app:CSnorm}
%%%%%%%%%%%%%%%%%%%%%%%%%%%%%%%%%%%%

We write the Chern-Simons action as follows:
\be
S_{\rm CS} = \frac{k}{4 \pi} \int_M \Tr' \left( \calA \wedge \d \calA + \frac{2}{3} \calA^3 \right) \ .
\ee
We define $\calA \equiv -i T_a A^a_\mu \d x^\mu$.  The $T_a$ are generators in some irreducible representation of a 
Lie algebra ${\mathfrak g}$ 
corresponding to a compact simple Lie group $G$.  The generators obey the commutation relations
$[T_a, T_b] = i {f_{ab}}^c T_c$.  
The field strength is defined to be $\calF = \d \calA + \calA^2$.  

We have defined the trace $\Tr'$ such that the action is independent of the choice of irreducible representation.  Let us review what this definition entails.  
Up to a constant of proportionality, all bilinear forms on $G$ are proportional to the Killing form.  
If we take an arbitrary representation of $G$ described by highest weight $\lambda$,
then we can choose the generators to be orthogonal
such that the standard trace satisfies the relation $\Tr ( T^a_\lambda T^b_\lambda) = |\theta|^2 x_\lambda \delta^{ab}$.  We have included the norm of the longest root $\theta$ because rescaling $T^a$ will rescale the length of all the weights and roots.  (Typically, one makes the choice $|\theta|^2 = 2$.)
The Dynkin index $x_\lambda$ should then be independent of this normalization.  
Taking a trace of both sides relates $x_\lambda$ to the quadratic Casimir from which one deduces that
\be
x_\lambda = \frac{\dim(\lambda)}{\dim({\mathfrak g})} \frac{(\lambda, \lambda + 2 \rho)}{|\theta|^2} \ ,
\ee
where $\rho = \frac{1}{2} \sum_{\alpha>0} \alpha$ is half the sum of the positive roots.
The normalized trace is defined such that ${\Tr}' = \Tr / |\theta|^2 x_\lambda$.

The representation independence of $\Tr'$ means we do not need to worry about the precise values of the Dynkin indices, but we give them for completeness.
For the fundamental representation of $SU(n)$ ($n\geq 2$) and $USp(2m)$ ($m\geq 1$), the Dynkin index is always $x_\lambda = 1/2$.  For the fundamental representation of $SO(n)$ ($n>3$), the Dynkin index is always $x_\lambda = 1$, while for $SO(3)$, $x_\lambda = 2$.   

We now review the gauge invariance 
argument that shows $k$ must be an integer.
Gauge invariance implies the path integral is invariant under gauge transformations,
\be
\calA \to g \calA g^{-1} + g \, \d g^{-1} \ , 
\ee
where $g \in G$.  
Up to boundary terms that we assume vanish, the variation of the action is the Wess-Zumino term
\be
\delta S_{\rm CS} = \frac{k}{12 \pi} \int_M \Tr' \left( \d g^{-1} \, g \right)^3 \ .
\ee
Naively, this term looks like it breaks gauge invariance, but in fact it is proportional to the
winding number $w$ of the map $g: M \to G$, i.e.\  $\delta S_{\rm CS} =2 \pi k \, w$.  
That $w$ is integer or half integer allows the path integral, which depends on $e^{i S_{\rm CS}}$, to be invariant if $k$ is chosen appropriately.

A straightforward calculation demonstrates that $w$ is integer for maps $g : S^3 \to SU(2)$.
Similarly, $w$ is half-integer for maps $g : S^3 \to SO(3)$.  We conclude that for $SU(2)$, $k$ must be an integer, while for $SO(3)$, $k$ must be an even integer. 

Note that $SU(2)$ is always a subgroup of $SU(n)$ (for $n\geq 2$) and $USp(2m)$ (for $m\geq 1$).  Also $SO(3)$ is always a subgroup of $SO(n)$ (for $n \geq 3$).  Thus, the maps $g : S^3 \to SU(2)$
and $g : S^3 \to SO(3)$ are also maps into the larger classical groups.
From this embedding and the representation independence of $\Tr'$, it follows that $k$ should be an integer for $SU$ and $USp$ groups and it should be an even integer for $SO$ groups.

%%%%%%%%%%%%%%%%%%%%%%%%%%%%%%%%%%%%
\section{Conventions for $\calN = 3$ SUSY Chern-Simons matter theories}
\label{app:N3}
%%%%%%%%%%%%%%%%%%%%%%%%%%%%%%%%%%%%

We review the construction of $\calN = 3$ SUSY CS matter theories.  Such theories have an $SU(2)_R$ R-symmetry while $\calN=2$ theories in contrast have only a $U(1)_R$.
We follow the notation of ref.\ \cite{arXiv:0704.3740}.
The CS action itself may be given $\calN=2$ SUSY by the addition of a fermion $\chi$ and two auxiliary scalars $D$ and $\sigma$:
\be
S_{\rm CS}^{\calN = 2} = \frac{k}{4\pi} \int \Tr' \left( \calA \wedge \d \calA + \frac{2}{3} \calA^3 \right)
- \frac{k}{4\pi} \int \Tr' \left( \overline \chi \chi - 2D \sigma \right) d^3 x \ .
\ee
A complete $\calN = 3$ action is then
\be
S = S_{\rm CS}^{\calN = 2} + \int d^3x \, d^4 \theta  \left(Q^\dagger e^V Q + \tilde Q^t e^{-V} \tilde Q^* \right)
+ \left[ \int d^3x \, d^2 \theta \left(- \frac{k}{4\pi} \Tr' \Phi^2 + \tilde Q \Phi Q \right) + c.c. \right] \ .
\ee
Here $Q$ and $\tilde Q$ are chiral superfields 
in the fundamental and anti-fundamental representation of the gauge group $G$ respectively.
The scalar components $q$ of $Q$ and $\tilde q^\dagger$ of $\tilde Q^\dagger$ 
fill out a doublet under $SU(2)_R$.  
The chiral field $\Phi$ is in the adjoint of $G$.  Its scalar component $\phi$ combines with $\sigma$
to give a three dimensional representation of  $SU(2)_R$.

We would like to verify some earlier claims about the global symmetry group under which the fundamental matter fields $Q$ and $\tilde Q$ transform when $G=SO(n)$ or $G=Sp(2m)$.
From appendix \ref{app:reps}, it is clear that 
the adjoint representation of $Sp(2m)$ is a symmetric tensor product of two fundamental representations.  
Similarly, the adjoint representation of $SO(n)$ is an antisymmetric tensor product. 
If we consider
$\Phi_{ab}$ where $a$ and $b$ are fundamental group indices, 
this property of the adjoint representation means that $\Phi_{ab}$ is a symmetric matrix for $Sp(2m)$ and an anti-symmetric matrix for $SO(n)$. 
We can take advantage of the symmetry properties of $\Phi$ to rewrite the superpotential as
\be
2 \tilde Q_a^j  Q_b^j \Phi_{ab} = (\tilde Q_a^j Q_b^j \pm Q_a^j \tilde Q_b^j) \Phi_{ab} \ .
\ee
where we use the plus sign for $G=Sp(2m)$ and the minus sign for $SO(n)$.  The indices 
$j,k = 1, \ldots, N_f$ index the flavors.  Summation over the indices is implied.

Let us start with $G=SO(n)$.  In this case, we can introduce a doublet field 
$X_a^I = (Q_a^j, \tilde Q_a^k)$, $I=1,\ldots, 2N_f$.  We also introduce the $2N_f \times 2N_f$
antisymmetric matrix
\be
J
=
\left(
\begin{array}{cccccc}
0 & {\rm Id} \\
-{\rm Id} & 0 
\end{array}
\right) \ .
\ee
With these two ingredients, the superpotential can be written in the form
\be
2 \tilde Q_a^j Q_b^j \Phi_{ab} = X^I_a J^{IJ} X^J_b  \Phi_{ab}\ .
\ee
This superpotential makes manifest a global $Sp(2N_f, \mathbb{C})$ symmetry.  The kinetic term
meanwhile preserves a $U(2N_f)$ global symmetry.   
In more detail, we can rewrite the kinetic term as
\be
Q^\dagger e^V Q + \tilde Q^t e^{-V} \tilde Q^* = Q^\dagger e^V Q + \tilde Q^\dagger (e^{-V})^t \tilde Q \ .
\ee
Given that the gauge group is $SO(2n)$, we know that $e^V$ 
is an orthogonal matrix and hence that $(e^{-V})^t = e^V$.  The kinetic term can thus be written in the form
$(X^I_a)^* (e^V)_{ab} X^I_b$.
The intersection of $USp(2N_f, \mathbb{C})$ and $U(2N_f)$ is $USp(2N_f)$.  

For $G=Sp(2m)$, we play a similar game.  First we need to recall that representations of $Sp(2m)$ are pseudoreal.  If $T_{ab}$ are Hermitian generators of the Lie algebra in the fundamental representation, we can write the pseudoreality condition as $\calJ^{-1} T \calJ = -T^* = - T^t$, where 
$\calJ_{ab}$ is the analog of $J$ acting on the color indices instead of the flavor indices. 
We introduce
the doublet field 
\be
X^I = ( Q^j + i \calJ \tilde Q^j, i Q^k +    \calJ \tilde Q^k) \ , 
\ee
We find that
\begin{eqnarray}
\frac{1}{2i} X_a^I X_b^I  (\calJ  \Phi)_{ab} &=& Q_a^j (\calJ \Phi)_{ab} (\calJ \tilde Q^j)_b
+ (\calJ \tilde Q^j)_a (\calJ  \Phi)_{ab} Q_b^j \nonumber \\
&=& 2 \tilde Q_a^j Q_b^j \Phi_{ab} \ .
\end{eqnarray}
where we have used the facts that $\calJ^2 = -1$ and that $\Phi = \Phi^t = \calJ \Phi \calJ$.  This last condition comes from the pseudoreality of representations of $Sp(2n)$.
This form of the superpotential makes manifest a 
global $O(2N_f, \mathbb{C})$ symmetry.  The kinetic term still preserves a $U(2N_f)$ symmetry although it's slightly more involved to see.  In more detail, note that
\begin{eqnarray*}
(X^I)^\dagger e^V X^I &=& 
(Q + i \calJ \tilde Q)^\dagger e^V (Q + i \calJ \tilde Q)
+  (iQ +  \calJ \tilde Q)^\dagger e^V (iQ +  \calJ \tilde Q) \\
&=& 2 Q^\dagger e^V Q +2 (\calJ \tilde Q)^\dagger e^V (\calJ \tilde Q) \\
&=& 2Q^\dagger e^V Q + 2 \tilde Q^\dagger (e^{-V})^t \tilde Q \ ,
\end{eqnarray*}
where in the last line we used the pseudoreality condition $\calJ e^V \calJ = -(e^{-V})^t$.
The intersection of $O(2N_f, \mathbb{C})$ and $U(2N_f)$ is $O(2N_f)$.

Something special can happen when the tensor product of the R-symmetry, gauge symmetry, and global symmetry representations is a real representation.  
Let us consider the hypermultiplet $(Q, \tilde Q^*)$.  
The R-symmetry representation is a pseudoreal doublet of $SU(2)$ where the analog of $J$ and $\calJ$ above is now $i \sigma_2$.   To get a real representation, the product of the gauge and global symmetries should be pseudoreal as well.  
Let us take the hypermultiplet to transform under the fundamental representation of $Sp(2m)$ and assume the flavor representation is real.  We consider the antilinear map $\tau$:
\be
\tau \left(
\begin{array}{c}
Q \\
\tilde Q^* 
\end{array}
\right)
= 
\calJ \otimes i \sigma_2 
\left(
\begin{array}{c}
Q^* \\
\tilde Q 
\end{array}
\right)
= 
\left(
\begin{array}{c}
\calJ \tilde Q \\
-\calJ Q^* 
\end{array}
\right) \ .
\ee
Note that $\tau^2 = 1$ and has eigenvalues $\pm 1$.  
A half hypermultiplet is an eigenvector of $\tau$.  
We can think of a half hypermultiplet as a full hypermultiplet with the constraint 
$Q = \pm \cal J \tilde Q$.

In the presence of half hypermultiplets, the argument demonstrating $O(2N_f)$ flavor symmetry needs to be adjusted.  We start with $N_f$ hypermultiplets which have a global $O(2N_f)$ symmetry.  Under the constraint $Q = \cal J \tilde Q$ leaving $N_f$ half hypermultiplets, only a $O(N_f)$ global symmetry is preserved.  Note that $N_f$ can be odd.

%%%%%%%%%%%%%%%%%%%
\bibliographystyle{ytphys}
\bibliography{orthosymplectic}
%%%%%%%%%%%%%%%%%%%

\end{document}